\documentclass[prd]{revtex4}

\usepackage{float}
\usepackage{graphicx}
\usepackage{caption}
\usepackage{amsmath}
\usepackage{array}
\usepackage{bm}
\usepackage{amssymb}
\usepackage{amsfonts}
\usepackage{amssymb}
\usepackage{color}

\begin{document}
\title{Multi-peak structure of meson spectral function in magnetic field}
\author{Haoran Li and Ziyue Wang}
\affiliation{School of Physics and Optoelectronic Engineering, Beijing University of Technology, Beijing 100021, China}
\date{\today}
\begin{abstract}
We investigate the spectral functions of neutral and charged mesons in a hot dense medium under a external magnetic field using the two-flavor quark-meson model within the functional renormalization group (FRG) framework. Our results show that the spectral functions of $\sigma$ and $\pi_0$ mesons develop new structures due to decay channels into quarks occupying different Landau levels. By consistently incorporating the momentum relations at vertices for charged particles in a magnetic field, we further show that the $\pi_+$ spectral function develops a multi-peak structure at finite temperatures, resulting from the various annihilation and decay channels available to $\pi_+$ in the magnetic environment. This multi-peak structure is further enhanced in a finite-density medium, causing the $\pi_+$ meson to become a broad resonance at lower temperatures and densities compared to neutral mesons. Such a multi-peak pattern is expected to be universal for charged mesons under magnetic fields and carries significant implications for understanding transport properties in magnetized strongly interacting fluids.\end{abstract}

\maketitle
\section{Introduction}
Quantum Chromodynamics (QCD) phase transitions at finite temperature and density provide important insights into strongly interacting matter, as realized in heavy-ion collisions and compact stars. Strong magnetic fields, which can reach up to $10^{19}$ Gauss ($|eB|\sim10m_\pi^2$) in the early universe, magnetars, and non-central heavy-ion collisions~\cite{Deng:2012aa,Skokov:2009aa,Grasso:2000wj}, can significantly modify QCD matter. At such field strengths, QED effects become comparable to QCD, leading to phenomena such as magnetic catalysis (MC)~\cite{Gusynin:aa,Gusynin:ab}, inverse magnetic catalysis (IMC)~\cite{Bruckmann:2013aa,Bruckmann:2013ab,Chao:2013aa}, as well as paramagnetic and diamagnetic responses~\cite{Hofmann:2021aa,Orlovsky:2014aa, Endrodi:2013aa,Fayazbakhsh:2014aa,Bali:2014aa}. In addition, electromagnetic fields induce non-equilibrium effects such as the chiral magnetic effect (CME)~\cite{Fukushima:2008aa,Kharzeev:2008aa,Kharzeev:2015aa}, highlighting the rich interplay between QCD and QED.

Meson properties in magnetic fields are essential for understanding both equilibrium and dynamical aspects of QCD matter. While meson spectra, reflecting static properties, have been extensively studied in lattice simulations and effective models\cite{Tripolt:2014aa,Dudek:2013aa,Blank:2011aa}, dynamical properties remain less explored. The spectral function provides a key probe of such dynamics, encoding information about excitation modes and decay processes. In particular, a single peak typically indicates a dominant quasiparticle mode, whereas multiple peaks signal the coexistence of different excitation channels and energy scales.

In recent years, continuum non-perturbative approaches such as Dyson–Schwinger equations (DSE) and functional renormalization group (FRG) have been widely used to investigate QCD matter under extreme conditions~\cite{Fischer:2011aa,Mueller:2014aa,Gao:2021aa, Fu:2020aa,Braun:2009gm,Kamikado:2013aa}. In this work, we employ the two-flavor quark-meson model combined with the FRG framework~\cite{Tetradis:2003aa,Herbst:2013aa,Fu:2015aa,Polonyi:2003aa,Wetterich:1992yh,Pawlowski:2005xe,Gies:2006wv}, which incorporates quantum and thermal fluctuations beyond the mean-field level and is well suited for studying chiral dynamics.

A key aspect of meson dynamics in a magnetic field is the modification of kinematics at interaction vertices. Unlike the case without magnetic field, where momentum conservation follows from plane-wave structure, the presence of Landau levels leads to nontrivial relations derived from coordinate-space integration of Landau eigenstates. By consistently incorporating these modified vertex relations into the FRG flow equations for two-point functions, we are able to resolve the microscopic processes contributing to meson spectral functions. Our main result is that the spectral function of the charged pion develops a pronounced multi-peak structure at finite temperature. This structure originates from a combination of decay channels $\pi_+'\to u_i\bar d_j$ and annihilation processes $\pi_+'\bar u_i\to \bar d_j$ and $\pi_+' d_i\to u_j$, involving quarks occupying different Landau levels. As temperature or density increases, these processes become more abundant, leading to a broad resonance structure of the $\pi^+$ meson at lower temperature and density compared to neutral mesons. The resulting multi-peak pattern reflects the underlying Landau-level-resolved dynamics and is expected to be a generic feature of charged mesons in magnetic fields, with potential implications for transport properties in magnetized strongly interacting matter.

This paper is organized as follows. In Sec.~\ref{s2}, we derive the FRG flow equations for the effective potential and meson two-point functions in the quark-meson model. In Sec.~\ref{s3}, we present the numerical procedure and results, including the phase diagram and meson spectral functions. We summarize our findings in Sec.~\ref{s4}. Technical details are provided in the appendices.

\section{Flow equations}
\label{s2}
The quark-meson model is a low-energy effective theory derived from the partial bosonization of four-fermion interaction models, while preserving the global symmetries of QCD. Its two-flavor version has been widely adopted as a useful framework for investigating chiral dynamics, as it captures both the spontaneous breaking of chiral symmetry in vacuum and its restoration at finite temperature and density\cite{Tetradis:2003aa,Herbst:2013aa,Fu:2015aa}. At energies below approximately $1$ GeV, the relevant mesonic degrees of freedom are the pseudoscalar ${\bm{\pi}}$ and the scalar $\sigma$ mesons, making this model particularly suitable for studying chiral aspects of QCD phenomenology. The Euclidean effective action of the model at finite temperature $T$ and under external magnetic field $\vec{B}=B\hat{z}$ is given as\cite{Andersen:2013aa}
\begin{equation}
\begin{split}
\Gamma_k&=\int_x\Big\{\bar{\psi}\Big[D\!\!\!\!/+h_k\left(\sigma+i\gamma_5\vec{\tau}\vec{\pi}\right)\Big]\psi+\frac{1}{2}(\partial_{\mu}\sigma)^2+\frac{1}{2}(\partial_{\mu}\pi_0)^2+\mathcal{D}_{\mu}\pi^+\mathcal{D}^{\mu}\pi^-+U_k(\phi^2)-c\sigma\Big\}\\
\end{split}
\end{equation}
with the meson field in $O(4)$ representation $\phi=(\sigma,\vec{\pi})$. The fermionic field $\psi$ are two flavor Dirac-spinors. $\vec{\tau}$ are the Pauli matrices in isospin space, the abbreviation stands for $\int_x=\int_0^{1/T}dx_0\int d^3x$. For this moment we neglect the scale dependence of Yukawa coupling and wave function renormalization, namely they are kept to be constant $Z=1$ (local potential approximation), and $h=3.2$ to fit the quark mass in vacuum. The explicit chiral symmetry breaking term $-c\sigma$ corresponds to a finite current quark mass. The interaction between external magnetic field and quark field is implemented through the covariant derivative
\begin{equation}
D\!\!\!\!/=\gamma^{\mu}D_{\mu},~~~~D_{\mu}=\partial_{\mu}-iQA_{\mu},~~~~Q=\text{diag}(+2e/3, -e/3).
\end{equation}
The interaction of charged pion with the magnetic field is included by minimum coupling $\mathcal{D}_{\mu}=\partial_{\mu}-ieA_{\mu}$, with Landau gauge $A=-By\delta_1^k$. 

\subsection{Flow equations of effective potential}
Quantum and thermal fluctuation are of particular importance in the vicinity of phase transitions and are conveniently included within the framework of the functional renormalization group(FRG). The core quantity in this approach is the averaged effective action $\Gamma_k$ at the RG scale $k$ in Euclidean space, its scale dependence is described by the flow equation \cite{Wetterich:1992yh}
\begin{equation}
\partial_k \Gamma_k =\text{Tr}\int_p\Big[\frac{1}{2}G_{\phi,k}(p)\partial_kR_{\phi,k}(p)-G_{\psi,k}(p)\partial_kR_{\psi,k}(p)\Big].
\label{flow}
\end{equation}
The flow equation (\ref{flow}) has the form of a one loop diagram. $G_{\phi,k}=(\Gamma_k^{(2)}[\phi]+R_{\phi,k})^{-1}$ and $G_{\psi,k}=(\Gamma_k^{(2)}[\psi]+R_{\psi,k})^{-1}$ are the FRG modified meson and quark propagator. $\Gamma_k^{(2)}[\phi]=\delta^2\Gamma_k/\delta\phi^2$ is second functional derivative with respect to bosonic field, and $\Gamma_k^{(2)}[\psi]=\delta^2\Gamma_k/\delta\psi\delta\bar\psi$ corresponds to fermionic field. The symbol $\text{Tr}$ represents the summation over all inner degrees of freedom of mesons and quarks. The flow of $\Gamma_k$ from UV to IR is controlled by the cutoff functions $R_k$. In order to suit the finite temperature and magnetic field background setting, we adopt a anisotropic regulator depending only on $p_3$ throughout this work \cite{Kamikado:2013aa,Wen:2023aa}, in the following we call it $p_3$-regulator. The bosonic and fermionic regulators are chosen to be
\begin{eqnarray}
\label{regulator}
R_{\phi,k}&=&p_3^{2}r_B(y),\nonumber\\
R_{\psi,k}&=&\gamma_3p_3r_F(y)
\end{eqnarray}
in momentum space, with $y=p_3^2/k^2$ and $r_B(y)=(1/y-1)\Theta(1-y)$ and $r_F(y)=(1/\sqrt y-1)\Theta(1-y)$. The regulators $R_{\phi,k}$ and $R_{\psi,k}$ in the propagators $G_\phi$ and $G_\psi$ amount to having regularized three-momenta $\tilde{p}_3^2={\bf p}^2(1+r_B(y))$ and $\tilde{p}_3={\bf p}(1+r_F(y))$ for bosons and fermions respectively. 

For the propagators $\pi^\pm$ and quarks in magnetic field, we take Landau-level representation \cite{Schwinger:1951nm,Miransky:2015aa}. The propagators in coordinate space can be written as the  Fourier transformation of the momentum space propagators,
\begin{equation}
\label{picpropagator}
G_{\pi^c,f}(r,r')=e^{i\Phi(\mathbf{r}_\perp,\mathbf{r}'_\perp)}\int\frac{d^4p}{(2\pi)^4}e^{-ip\cdot(r-r')}{G}_{\pi^c,f}(p),
\end{equation}
the subscript $\pi^c=\pi^\pm$ for charged pion and $f=u,d$ for quarks. $\Phi(\mathbf{r}_\perp,\mathbf{r}'_\perp)$ is the Schwinger phase defined by $\Phi(\mathbf{r}_\perp,\mathbf{r}'_\perp)=-\frac{s_\perp(x-x')(y+y')|qB|}{2}$, where $s_\perp=\text{sign}(qB)$, $q$ is the charge of the charged complex scalar field or the fermion field. ${G}_{\pi^c,f}(p)$ is the corresponding propagators in momentum space. For charged complex scalar field, 
\begin{equation}
{G}_{\pi^c}(p)~=~2ie^{-p^2_\perp l^2}\sum_{n=0}^{\infty}\frac{(-)^nL_n(2p^2_\perp l^2)}{p_0^2-(2n+1)|eB|-(p^3)^2-m^2},
\end{equation}
where $L_n$ is the Laguerrel polynomial, and $p_\perp^2=p_x^2+p_y^2$, $l^2=1/|eB|$. For quarks, 
\begin{eqnarray}
\label{udpropagator}
G_f(p)&=&ie^{-p^2_\perp l_f^2}\sum_{n=0}^{\infty}(-1)^n\frac{D_{f,n}(p)}{(p_0+\mu)^2-p_3^2-2n|q_fB|-m^2},\nonumber\\
D_{f,n}(p)&=&{2[(p_0+\mu)\gamma^0-p^3\gamma^3+m][\mathcal{P}_+L_n(2p^2_\perp l_f^2)-\mathcal{P}_-L_{n-1}(2p^2_\perp l_f^2)]+4\vec{\gamma}_\perp\cdot \vec{p}_\perp L_{n-1}^1(2p_\perp^2l_f^2)},
\end{eqnarray}
where $\mathcal{P}_\pm=\frac{1}{2}[1\pm i\gamma^1\gamma^2 s_\perp^f]$ is the projector. For u-quark $s_\perp^u=+1$, $l_u^2=\frac{3}{2}l^2$, and for d-quark $s_\perp^d=-1$, $l_d^2=3l^2$. 

The flow equation of the effective potential is
\begin{eqnarray}
\label{flowU}
\partial_kU_k&=&\frac{1}{2}I_\sigma^{(1)}+\frac{1}{2}I_{\pi^0}^{(1)}+\frac{1}{2}\widetilde{I}_{\pi^+}^{(1)}+\frac{1}{2}\widetilde{I}_{\pi^-}^{(1)}-N_c\sum_{f=u,d}I_{f}^{(1)},
\end{eqnarray}
where $I_\phi^{(1)}=\frac{k^2}{\pi^2}\int_0^{\infty}p_{\perp}dp_{\perp}I_\phi^{(1)T}(E_\phi)$,  $\widetilde{I}_\phi^{(1)}=\frac{k^2|eB|}{\pi^2}\sum_{n=0}^{\infty}I_\phi^{(1)T}(E_\phi)$ and $I_{f}^{(1)}=\frac{k^2|q_fB|}{\pi^2}\sum_{n=0}^{\infty}\alpha_n I_f^{(1)T}(E_f)$ are the loop functions encoding the momentum integral and Matsubara sum in the neutral meson loop, charged pion loop, and quark loop respectively, $\alpha_n=1$ for $n=0$, and $\alpha_n=2$ for $n\geq1$. $I_\phi^{(1)T}(E_\phi)$ and $I_f^{(1)T}(E_f)$ are the threshold function encoding Matsubara sum, they presented in Appendix \ref{threshold}. The scale-dependent effective potential $U_k$ is a function of the invariant $\rho=\phi^2=\sigma^2+\vec{\pi}^2$. The curvature mass of the mesons and the quark mass are determined from 
\begin{equation}
\label{squaremass}
m^2_{\pi}=2U',\qquad\qquad m^2_{\sigma}=2U'+4\rho U'',\qquad\qquad m^2_{q}=h^2\rho,
\end{equation}
where the derivatives are $U'=\partial_\rho U$ and $U''=\partial^2_\rho U$, the curvature at the minimum of the effective potential yield the squared curvature masses of the pion and sigma meson. The energy of neutral meson, charged meson and quark are given correspondingly 
\begin{equation}
\label{energy}
E_{\phi}=\sqrt{k^2+p_\perp^2+m_{\phi}^2},\qquad 
E_{\pi^\pm}=\sqrt{k^2+(2n+1)|eB|+m^2_{\pi}},\qquad 
E_{f}=\sqrt{k^2+2n|q_fB|+m_{q}^2},
\end{equation}
where $p_\perp$ is the perpendicular momentum, $n$ is the Landau level of ${\pi^\pm}$ and quark. When turning off the magnetic field, one can simply replacing $(2n+1)|eB|\rightarrow p_\perp^2$ and $2n|q_fB|\rightarrow p_\perp^2$ in corresponding energy of $\pi^\pm$ and quark. The threshold function will return to those without magnetic field by taking the replacement $|eB|\sum_{n=0}^{\infty}\rightarrow\int p_{\perp}dp_{\perp}$ in $\widetilde{I}_{\pi^\pm}^{(1)}$ and $|q_fB|\sum_{n=0}^{\infty}\alpha_n\rightarrow 2\int p_{\perp}dp_{\perp}$ in $I_{f}^{(1)}$. 

\subsection{Flow equation of 2-point function}
We now consider the flow equations for meson two-point functions. From the flow equation for the effective action, we get only the statical properties of the mesons, namely the curvature masses defined through the effective potential. However, the thermal medium and magnetic field have much more complicated effects on the dynamical properties of mesons. From the two-point functions, one can extract the spectral functions which contain the information on full pole masses and decay properties of the mesons. In the following we consider the flow equations for two-point functions $\Gamma_{k,p}^{(2)}[\phi]$ at vanishing external momentum ${\bf p}=0$ for neutral mesons and lowest Landau level for $\pi^\pm$.

The flow equation of two-point function is derived from the flow equation for the effective action $\Gamma_k$ by taking second order functional derivative with respect to $\phi_i$ and $\phi_i^\dag$,
\begin{eqnarray}
\label{two-point}
\partial_k\Gamma^{(2)}_{k,p}[\phi_i] &=& \widetilde{\partial}_k\text{Tr}\int_q \Big[{1\over 2}G_{\phi,k}(q)\Gamma_k^{(4)}[\phi,\phi_i]-\frac{1}{2}G_{\phi,k}(q)\Gamma_k^{(3)}[\phi,\phi_i]G_{\phi,k}(q+p)\Gamma_k^{(3)}[\phi,\phi_i]\nonumber\\
&&\ \ \ \ \ \ \ \ \ \ \ \ +G_{\psi,k}(q)\Gamma_k^{(3)}[\psi,\phi_i]G_{\psi,k}(q+p)\Gamma_k^{(3)}[\psi,\phi_i]\Big],
\end{eqnarray}
where the symbol $\widetilde\partial_k$ means the derivative only on the regulators $R_k$. It is worth emphasizing that in the above expression, the conservation of momentum is taken as an underlying assumption. However, in the background magnetic field, the translation invariance no longer exists, neither the conservation of momentum. The evaluation of the above flow equation hence need to be started from the coordinate space. 

From the definition $\Gamma^{(3)}[\phi,\phi_i]=\delta\Gamma^{(2)}[\phi]/\delta\phi_i$, $\Gamma^{(3)}[\psi,\phi_i]=\delta \Gamma^{(2)}[\psi]/\delta\phi_i$ and $\Gamma^{(4)}[\phi,\phi_i]=\delta^2 \Gamma^{(2)}[\phi]/\delta\phi_i^2$ with $\phi_i$ the external field, we have the coupling matrices. The matrix elements are defined by $\Gamma^{(3)}_{ijm}=\partial^3\Gamma_k/\partial\phi_i\partial\phi_j\partial\phi_m$ and $\Gamma^{(4)}_{ijmn}=\partial^4\Gamma_k/\partial\phi_i\partial\phi_j\partial\phi_m\partial\phi_n$. The non-zero matrix components are $\Gamma^{(3)}_{\sigma\sigma\sigma}=12\sigma_kU''_k+8\sigma_k^3U^{(3)}_k$, $\Gamma^{(3)}_{\sigma\pi\pi}=4\sigma_kU''_k$, $\Gamma^{(4)}_{\sigma\sigma\sigma\sigma}=12U''_k+48\sigma_k^2U^{(3)}_k+16\sigma_k^4U^{(4)}_k$, $\Gamma^{(4)}_{\sigma\sigma\pi\pi}=4U''_k+8\sigma_k^2U^{(3)}_k$, $\Gamma^{(4)}_{\pi_0\pi_0\pi_0\pi_0}=12U''_k$, $\Gamma^{(4)}_{\pi_+\pi_-\pi_+\pi_-}=8U''_k$, $\Gamma^{(4)}_{\pi_+\pi_-\pi_0\pi_0}=4U''_k$. The derivatives are $U'=\delta U/\delta \rho$, $U''=\delta^2 U/\delta \rho^2$, $U^n=\delta^n U/\delta \rho^n$, and $\sigma_k$ is the sigma condensate at certain scale $k$. The vertex related to quarks are $\Gamma^{(3)}_{\bar{\psi}\sigma\psi}=g\times\mathbf{1}_D\times\mathbf{1}_f$, $\Gamma^{(3)}_{\bar{\psi}\pi_+\psi}=ig\gamma_5\sigma_+$, $\Gamma^{(3)}_{\bar{\psi}\pi_-\psi}=ig\gamma_5\sigma_-$, $\Gamma^{(3)}_{\bar{\psi}\pi_0\psi}=ig\gamma_5\sigma_3$. 

The loops in $\sigma$ and  $\pi^0$ two-point function are relatively simple, with the external field neutral, the loops are constructed by particles with equal but opposite charges, hence the Schwinger phase in the propagators can cancel out. In the following we present the flow equations for the pion and sigma meson 2-point functions in the external magnetic field
\begin{eqnarray}
\label{twopoint_neutral}
\partial_k\Gamma_{k,p_0}^{(2)}[\sigma]
&=&(\Gamma^{(3)}_{\sigma\sigma\sigma})^2J_{\sigma\sigma}(p_0)
+(\Gamma^{(3)}_{\sigma\pi\pi})^2\Big( J_{\pi_0\pi_0}(p_0)
+\widetilde{J}_{\pi_+\pi_+}(p_0)+\widetilde{J}_{\pi_-\pi_-}(p_0)\Big)\nonumber\\
&&-\frac{1}{2}\Gamma^{(4)}_{\sigma\sigma\sigma\sigma} I^{(2)}_{\sigma}-\frac{1}{2}\Gamma^{(4)}_{\sigma\sigma\pi\pi}(I^{(2)}_{\pi^0}+\widetilde{I}^{(2)}_{\pi^+}+\widetilde{I}^{(2)}_{\pi^-})\nonumber\\
&&+2g^2N_c( J^\sigma_{uu}(p_0,\mu)+J^\sigma_{dd}(p_0,\mu)),\nonumber\\
\partial_k\Gamma_{k,p_0}^{(2)}[\pi^0]
&=&
(\Gamma^{(3)}_{\sigma\pi\pi})^2 \Big(J_{\sigma\pi_0}(p_0)+J_{\pi_0\sigma}(p_0)\Big)
\nonumber\\
&&-\frac{1}{2}\Gamma^{(4)}_{\sigma\sigma\pi\pi} I^{(2)}_{\sigma}
-\frac{1}{2}\Gamma^{(4)}_{\pi\pi\pi\pi} I^{(2)}_{\pi_0}
-\frac{1}{2}\Gamma^{(4)}_{\pi\pi\tilde{\pi}\tilde{\pi}}(\widetilde{I}^{(2)}_{\pi^+}+\widetilde{I}^{(2)}_{\pi^-})\nonumber\\
&&+2g^2N_c(J_{uu}^{\pi_0}(p_0,\mu)+J_{dd}^{\pi_0}(p_0,\mu)).
\end{eqnarray}
In this formalism, the loop functions are defined as follows. $J_{\alpha\alpha}(p_0)=\frac{k^2}{\pi^2}\int_0^\infty q_\perp dq_\perp J_{\alpha\alpha}^T$ and $J_{\alpha\beta}(p_0)=\frac{k^2}{\pi^2}\int_0^\infty q_\perp dq_\perp J_{\alpha\beta}^T$ represent loop integrals with three-line vertices, where the former involves identical neutral mesons and the latter different neutral mesons. For charged mesons, the corresponding loop function is given by $\widetilde{J}_{\alpha\alpha}(p_0)=\frac{k^2|eB|}{\pi^2}\sum_{l=0}^{\infty} J_{\alpha\alpha}^T$ which incorporates summation over Landau levels. Similarly, $I^{(2)}_{\phi}=\frac{k^2}{\pi^2}\int_0^\infty q_\perp dq_\perp I^{(2)T}_{\phi}$ and $\widetilde{ I}^{(2)}_{\phi}=\frac{k^2|eB|}{\pi^2}\sum_{l=0}^{\infty}  I^{(2)T}_{\phi}$ denote loop functions with four-line vertices for neutral and charged mesons, respectively. For quark loops, the function $J^\phi_{ff}(p_0,\mu)=\frac{k^2|q_fB|}{\pi^2}\sum_{n}^{\infty}\alpha_n{J}_{ff}^{\phi T}(p_0,\mu)$ is introduced, where $f=u,d$ labels the quark flavor and $\phi=\sigma,\pi^0$ indicates the external meson line. The Matsubara summation, which are encapsulated in the threshold functions $J_{\alpha\alpha}^T$,$J_{\alpha\beta}^T$, $I^{(2)T}_{\phi}$ and ${J}_{ff}^{\phi T}$, are presented in Appendix~\ref{threshold}.

In order to investigate the medium effect on the meson mass and decay properties, we focus on spectral function in the particle rest frame, with vanishing spatial momentum $\vec p=0$. However, for charged pion in the external magnetic field, this conventional approach is not directly applicable due to the zero-point energy associated with Landau levels. Instead, we consider the ground Landau level $n=0$ with longitudinal momentum $p_3=0$ for the external line. The absence of translational invariance in charged particle propagators under a magnetic field requires that loop calculations be performed directly in coordinate space. Specifically, the flow equation for the two-point function of $\pi_\pm$ includes loops formed by $\pi_\pm$ and $\sigma$ mesons, as well as quark loops involving $u$- and $d$-quarks—all of which carry different electric charges. In such configurations, the Schwinger phase does not cancel, and standard momentum conservation at vertices is modified. Consequently, these loops must be evaluated starting from their coordinate-space representation. For a general external $\pi_\pm$ line in Landau level $n$ with momentum components $p_0$ and $p_3$, the corresponding eigenstate is given by $\phi_{np}({r})=\langle r|np\rangle=e^{-ip_0 t+ip_3z}\Psi_{np_1}(\vec{r}_\perp)$, which remains a plane wave along the $t$- and $z$-directions, while in the transverse plane it is described by a Landau-level wave function. Explicitly,
\begin{eqnarray}
\label{externalline}
\phi_{np}({r})=e^{-ip_0 t+ip_3z}\frac{e^{-\frac{\left(\frac{y}{l}+s_\perp p_1 l\right)^2}{2}}}{\sqrt{2^nn!\sqrt{\pi}l\;}}H_n\left(\frac{y}{l}+s_\perp p_1 l\right)e^{ip_1x},
\end{eqnarray}
where $l=1/\sqrt{qB}$ is the magnetic length, and $l=1/\sqrt{eB}$ for $\pi_\pm$. 
A meson loop composed by $\sigma$ and $\pi_+$ with two three-line vertex is calculated starting from a coordinate space loop
\begin{eqnarray}
\text{M-Loop}&=&(\Gamma^{(3)}_{\sigma\pi\pi})^2\int_{r,r'}\phi_{np}({r'})G_{\pi_+}(r,r')G_{\sigma}(r',r)\phi^*_{np}({r}).
\end{eqnarray}
Likewise, we also need to evaluate the quark loop constructed by u quark and d quark starting from the coordinate space. The quark propagator in the coordinate space can also be obtained by Fourier transformation \eqref{picpropagator} from the momentum space propagator \eqref{udpropagator}. The quark loop in coordinate space reads
\begin{eqnarray}
\text{F-loop}&=&\int_{r,r'}\text{Tr}\Big[\phi_{np}({r'})(\sqrt{2}ig\gamma_5)G_u(r,r')(\sqrt{2}ig\gamma_5)G_d(r',r)\phi^*_{np}({r})\Big]. 
\end{eqnarray}
The detailed calculation of the M-loop and F-loop are contained in the Appendix \ref{MloopFloop}. After complicated calculation of integrating out the coordinate and momentum, as well as considering the FRG modification to the loop and taking the derivative to the RG-scale, we finally arrive at the flow equation of the $\pi^+$ two point function, 
\begin{eqnarray}
\partial_k\Gamma_{\pi^+,k}^{(2)}
&=&
(\Gamma^{(3)}_{\sigma\pi\pi})^2\Big(\overline{J}_{\sigma\pi_c}(p_0)+\overline{J}_{\pi_c\sigma}(p_0)\Big)
\nonumber\\
&&-\frac{1}{2}\Gamma^{(4)}_{\sigma\sigma\pi\pi}I^{(2)}_{\sigma}
-\frac{1}{2}\Gamma^{(4)}_{\pi\pi\tilde{\pi}\tilde{\pi}} I^{(2)}_{\pi_0}
-\frac{1}{2}\Gamma^{(4)}_{+-+-}(\widetilde{I}^{(2)}_{\pi_+}+\widetilde{I}^{(2)}_{\pi_-})\nonumber\\
&&+2g^2N_c\Big(J^{\pi_+}_{du}(p_0,\mu)+J^{\pi_+}_{ud}(-p_0,\mu)\Big),
\end{eqnarray}
where $\overline{J}_{\sigma\pi_c}(p_0)$ is the loop function of a loop with three-line vertex composed by $\sigma$ and $\pi^+$. It is defined by 
\begin{eqnarray}
\overline{J}_{\sigma\pi_c}(p_0)=\frac{k^2}{\pi^2}\sum_{m=0}^{\infty}\int q_\perp dq_\perp\frac{1}{m!}\Big(\frac{q_\perp^2}{2|eB|}\Big)^m e^{-\frac{q_\perp^2}{2|eB|}} J_{\sigma\pi_c}^T(p_0),
\end{eqnarray}
where $J_{\sigma\pi_c}^T(p_0)$ encodes the Matsubara sum, which is defined in Appendix \ref{threshold}. Energy of $\sigma$ meson and $\pi^+$ meson are given by \eqref{energy}. $J^{\pi_+}_{du}(p_0,\mu)$ and $J^{\pi_+}_{ud}(-p_0,\mu)$ are the loop function of a quark loop composed by u and d quark, 
\begin{eqnarray}
\label{udloop}
J^{\pi_+}_{du}(p_0,\mu)&=&-\frac{2k^2|eB|}{3\pi^2} \sum_{n,m=0}^\infty  \Big\{Y_1 J_{du,mn}^{(0)T}(p_0,\mu)+\Big(4m|q_dB|Y_1-\frac{16}{3}|eB| Y_2\Big)J_{du,mn}^{(1)T}(p_0,\mu)\Big\},\nonumber\\
J^{\pi_+}_{ud}(p_0,\mu)&=&-\frac{2k^2|eB|}{3\pi^2} \sum_{n,m=0}^\infty  \Big\{Y_1 J_{ud,nm}^{(0)T}(p_0,\mu)+\Big(4n|q_uB|Y_1-\frac{16}{3}|eB| Y_2\Big)J_{ud,nm}^{(1)T}(p_0,\mu)\Big\}.
\end{eqnarray}
$J_{du,mn}^{(0)T}$, $J_{du,mn}^{(1)T}$ and $J_{ud,nm}^{(0)T}$, $J_{ud,nm}^{(1)T}$ encodes the Matsubara sum in the loop, where $m$ is the Landau level in the propagator of d-quark, and $n$ is the Landau level in the propagator of u-quark. The details of threshold functions are included in the appendix.\ref{threshold}. The function $Y_1$ and $Y_2$ encodes information in perpendicular momentum integral, 
\begin{eqnarray}
\label{Y1Y2}
Y_1(m,n)&\equiv&\frac{(-1)^{m}}{5^{n+1}}\int du e^{-\frac{3}{5}u}\Big[L_n(-\frac{2}{5}u)L_{m-1}(u)
 -5L_{n-1}(-\frac{2}{5}u)L_{m}(u)\Big]
 =-\frac{2^{m-1}}{3^{n+m}}\frac{(2n+m)\Gamma(n+m)}{\Gamma(1+n)\Gamma(1+m)},\nonumber\\
Y_2(m,n)&\equiv&\frac{(-1)^{m}}{5^{n+1}}\int du e^{-\frac{3}{5}u} u L_{n-1}^1(-\frac{2}{5}u)L_{m-1}^1(u)
=-\frac{ 2^{m-1}}{3^{n+m}}\frac{\Gamma(n+m)}{\Gamma(n)\Gamma(m)}.
\end{eqnarray}
These integrals are performed using
\begin{eqnarray}
\int_0^{+\infty}e^{-bx}x^\alpha L_n^\alpha(\lambda x)L_m^\alpha(\mu x) dx=\frac{\Gamma(m+n+\alpha+1)}{m!n!}\frac{(b-\lambda)^n(b-\mu)^m}{b^{m+n+\alpha+1}}F[-m,-n;-m-n-\alpha,\frac{b(b-\lambda-\mu)}{(b-\lambda)(b-\mu)}],
\end{eqnarray}
where F is the hypergeometric function. In deriving the quark propagator, the condition $L_{-1}(x)=0$ is assumed. As a result, the analytic expression for$Y_1(m,n)$ is valid when $m>0$ or $n>0$, while $Y_1(0,0)=0$. Similarly, the definition of $Y_2(m,n)$ implies that it is nonzero only when $m\geq 1$ and $n\geq 1$. Consequently, the quark loop contribution in the flow equation vanishes when both quarks in the propagators occupy the lowest Landau level, but becomes nonvanishing for any other combination of Landau levels. Physically, this is because a spin-0 $\pi_+$ meson cannot decay into a pair of $u\bar{d}$-quarks both at ground Landau level whose spins are parallel to the magnetic field. The functions $Y_1$ and $Y_2$ exhibit a structure reminiscent of a binomial distribution. Specifically, for a d-quark in the $m$-th Landau level and a u-quark in the $n$-th Landau level, the amplitude is weighted by a factor proportional to $\frac{(n+m-1)!}{(n-1)!m!}(\frac{2}{3})^m(\frac{1}{3})^{n-1}$. This specific relation between momentum and Land au levels at the $\pi_+$-quark vertices gives rise to diverse annihilation channels for $\pi_+$ in a thermal medium. As will be shown in the following section, these channels are responsible for the multi-peak structure observed in the $\pi_+$ spectral function at finite temperature.

In order to obtain the two-point functions in Minkowski space, we perform the analytic continuation
\begin{equation}
\Gamma^{(2)}_{k,\omega}[\phi_i] = \lim_{\epsilon\to 0}\lim_{p_0\to -i(\omega+i\epsilon)}\Gamma^{(2)}_{k,p_0}[\phi_i].
\end{equation}
This substitution of the discrete Euclidean frequency $p_0$ by the continuous energy $\omega$ is done explicitly before the integration of the RG scale $k$. Finally, the meson spectral functions are expressed in terms of the imaginary and real parts of the retarded propagator,
\begin{equation}
\rho_{k,\omega}[\phi_i]=-\frac{1}{\pi}\frac{\text{Im}\Gamma^{(2)}_{k,\omega}[\phi_i]}{\left[\text{Re}\Gamma^{(2)}_{k,\omega}[\phi_i]\right]^2+\left[\text{Im}\Gamma^{(2)}_{k,\omega}[\phi_i]\right]^2}.
\end{equation}

\section{Numerical process and Results}
\label{s3}
The present study is performed within the Local Potential Approximation (LPA), where wave-function renormalization and momentum-dependent interactions are neglected. Within this truncation, the FRG framework is expected to capture the qualitative phase structure in the temperature-driven crossover region and at small to moderate chemical potentials, as well as the threshold structure of spectral functions and the kinematics associated with Landau-level quantization. However, the LPA does not include momentum-dependent dressing effects, nor the scale dependence of wave-function renormalization. In particular, anisotropic effects in the presence of a magnetic field, such as distinct renormalization factors $Z_\perp$ and $Z_\parallel$, are not taken into account. At larger chemical potentials, derivative operators and wave-function renormalization effects may qualitatively modify the phase structure and alter the relation between curvature masses and pole masses \cite{Fu:2020aa,Yin:2019aa, Fu:2022gou}. As a consequence, quantitative features such as the precise peak widths and pole positions may be affected by the truncation. The results presented here should therefore be understood as providing a qualitative description of the spectral structure, in particular the emergence of multi-peak patterns and their relation to Landau-level-resolved processes, rather than a precision determination of meson properties.

In order to investigate spectral functions at finite temperature and density, we employ both the Taylor expansion method~\cite{Pawlowski:2014aa,Stokic:2010aa} and the grid method~\cite{Schaefer:2005aa} to solve the flow equation \eqref{flowU}. In the Taylor expansion method, the effective potential is expanded around a fixed expansion point,
\begin{equation}
U_k(\rho) =
\sum_{n=0}^{n_{\max}} \frac{\lambda_{n,k}}{n!} (\rho - \rho_0)^n,
\end{equation}
where $\rho_0$ is chosen as the minimum of the effective potential at $k = 0$, and the expansion is truncated at $n_{\max} = 5$. Since the Taylor expansion may not fully capture the behavior across first-order phase transitions, we also employ the grid method to obtain a more complete description.

In the grid method, the field variable $\rho$ is discretized on a uniform grid with 200 points in the range $[0, \rho_{\max}]$, with $\rho_{\max} = 200^2~\text{MeV}^2$. During the RG evolution from the ultraviolet to the infrared scale, the condensate is determined by locating the minimum of the scale-dependent effective potential $U_k$. Compared to the conventional three-momentum regulator, the $p_3$-regulator used in this work suppresses fluctuations only along the longitudinal direction. As a result, the flow needs to be evolved to sufficiently small infrared scales, and the effective potential may develop enhanced concavity in the small-$\rho$ region, where $U''$ becomes negative and the sigma mass squared may turn negative. To ensure numerical stability, we implement two alternative treatments. Method A: whenever $m_\sigma^2$ becomes negative, it is set to zero in the affected region to allow the flow to continue. Method B: once $m_\sigma^2$ becomes negative in an interval $[0, \rho_i]$ at a given scale, the evolution is restricted to the reduced domain $[\rho_i, \rho_{\max}]$ for all subsequent scales. We have verified that both approaches, as well as the Taylor expansion method, yield consistent results away from the critical endpoint. Close to the critical endpoint, the Taylor expansion becomes unreliable, while grid method B remains applicable. In practice, grid method A is found to be robust across the full phase diagram.

With the anisotropic $p_3$-regulator, the transverse momentum integrals are not explicitly UV regulated, which leads to a regulator-induced divergence in the flow equation of the effective potential \eqref{flowU}. In the Taylor expansion method, this divergence does not affect the flow of the expansion coefficients, since it is independent of the field variable $\rho$ and drops out after taking derivatives. 

In the grid method, we remove this field-independent contribution by evolving a subtracted effective potential,
\begin{equation}
\Delta U_k(\rho) \equiv U_k(\rho) - U_k(\rho_{\rm ref}),
\end{equation}
where $\rho_{\rm ref}$ is chosen as the minimum of the effective potential. The corresponding flow equation is
\begin{equation}
\partial_k \Delta U_k(\rho)
= \partial_k U_k(\rho) - \partial_k U_k(\rho_{\rm ref}).
\end{equation}
Since the regulator-induced divergence originates from the large transverse momentum region and is independent of $\rho$, it cancels identically in the subtracted flow. Therefore, the quantity $\Delta U_k(\rho)$ evolved in the grid method is finite and well-defined.

The numerical solution requires specification of the model parameters and initial conditions. In the ultraviolet, we choose
\begin{equation}
U_\Lambda = \frac{1}{2} m_\Lambda^2 \rho + \frac{1}{4} \lambda_\Lambda \rho^2.
\end{equation}
The parameters are fixed by matching vacuum observables at $k = 0$. Using $m_q = 300~\text{MeV}$, $m_\pi = 137~\text{MeV}$ and $f_\pi = 93~\text{MeV}$, we obtain the corresponding initial conditions as specified below. The transverse momentum integration is performed up to $p_T^{\max} = 5~\text{GeV}$, and the Landau-level summation is truncated at an energy cutoff of 5 GeV.

For the two-point functions, we use the standard initial conditions
\begin{equation}
\Gamma^{(2)}_{\Lambda,\omega}[\sigma] = -\omega^2 + 2U'_\Lambda + 4\rho U''_\Lambda,\quad
\Gamma^{(2)}_{\Lambda,\omega}[\pi^0] = -\omega^2 + 2U'_\Lambda,\quad
\Gamma^{(2)}_{\Lambda,\omega}[\pi^+] = -\omega^2 + |eB| + 2U'_\Lambda.
\end{equation}
In the analytic continuation, we introduce a small imaginary part $\epsilon = 1~\text{MeV}$, which is kept fixed throughout the calculation.

\subsection{Phase diagram and screening masses}
\begin{figure}[H]
\centering
\includegraphics[height=0.25\textwidth]{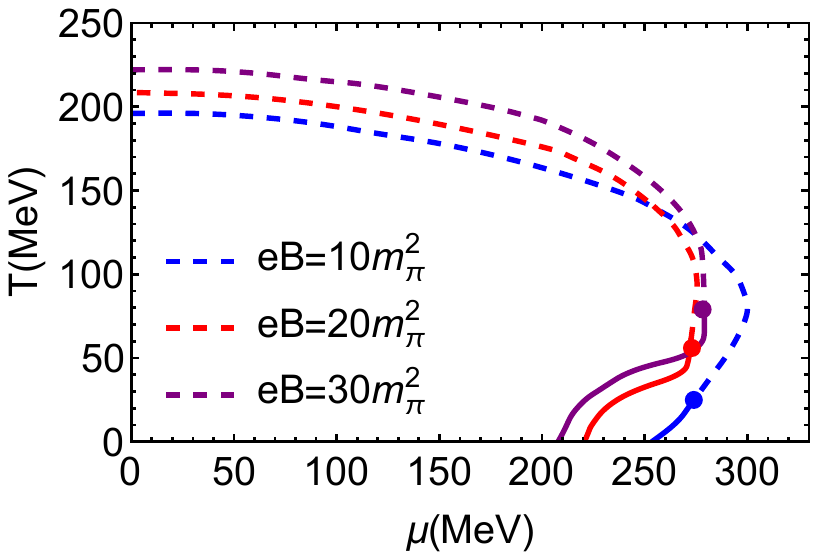}
\caption{Phase boundaries in the $T-\mu$ plane under different magnetic fields. Dashed lines denote crossovers; solid lines indicate first-order transitions, circles mark the critical endpoints.}
\label{phase_boundary}
\end{figure}

We begin with a brief overview of the phase diagram, which provides essential input for the subsequent calculation of two-point correlation functions. As shown in Fig.\ref{phase_boundary}, the phase structure is consistent with typical results of the quark-meson model. In agreement with Ref. \cite{Kamikado:2013aa}, the system exhibits magnetic catalysis, with the chiral restoration temperature increasing as the magnetic field strength grows. 
At the same time, the first-order phase transition becomes more pronounced with increasing magnetic field, extending to lower chemical potentials and persisting up to higher temperatures. Consequently, the critical endpoint (CEP) shifts toward higher temperatures. The CEP locations are found at $(T,\mu) = (25,274)\,\text{MeV}$, $(56,272)\,\text{MeV}$, and $(79,278)\,\text{MeV}$ for $eB = 10, 20, 30\,m_\pi^2$, respectively.

To assess the stability of the phase structure, we have examined the dependence on the transverse cutoff $\Lambda_T$ and the grid resolution. The results are summarized in Appendix \ref{stability_analysis}. While quantitative details such as the precise location of the phase boundary show some sensitivity to these parameters, the overall topology of the phase diagram remains unchanged. Since the main focus of this work is on meson spectral functions rather than precision determination of the phase boundary, this level of stability is sufficient for our purposes.

In what follows, we present spectral functions at various temperatures for $eB = 10\,m_\pi^2$, where a large number of annihilation and decay channels appear within the relevant energy range. We also analyze the chemical-potential dependence along a vertical trajectory crossing the CEP. For $eB = 20\,m_\pi^2$, the critical temperature $T_c = 56\,\text{MeV}$ is sufficiently high for thermal effects to become visible in the spectral functions. For even larger magnetic fields, although $T_c$ further increases, the corresponding decay thresholds move to higher energies beyond the region of interest. 

\begin{figure}[H]
\centering
\includegraphics[height=0.25\textwidth]{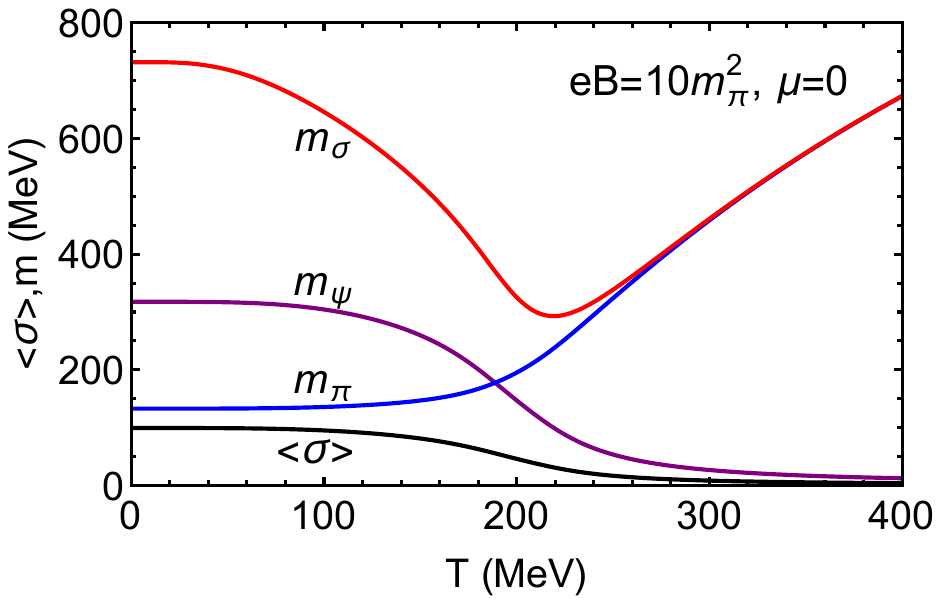}
\includegraphics[height=0.25\textwidth]{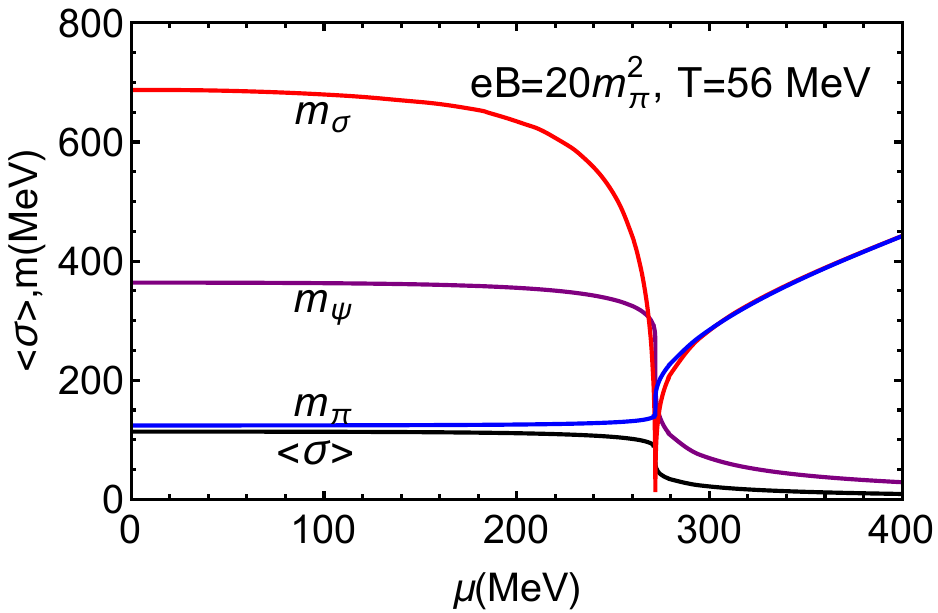}
\caption{Left: temperature dependence of meson screening masses $m_\sigma$, $m_\pi$, quark mass $m_\psi$, and the order parameter $\sigma$ at $\mu = 0$ and $eB = 10 m_\pi^2$. Right: chemical potential dependence of meson screening masses $m_\sigma$, $m_\pi$, quark mass $m_\psi$, and the order parameter $\sigma$ at $T = 56$ MeV and $eB = 20 m_\pi^2$.}
\label{condensate}
\end{figure}

To better identify the relevant energy scales entering the spectral functions, we therefore examine the temperature and chemical-potential dependence of meson screening masses, which provide a useful guide to the kinematic thresholds. The left panel of Fig. \ref{condensate} shows the meson screening masses $m_\sigma$ and $m_\pi$, the quark mass $m_\psi$, and the order parameter $\sigma$ as functions of temperature at zero chemical potential under $eB = 10\ m_\pi^2$. A common way to define the pseudo-critical temperature for the crossover is to locate the point where the order parameter varies most rapidly, giving $T_c = 196$ MeV for $eB = 10\ m_\pi^2$. At high temperatures, the sigma and pion masses become degenerate, reflecting the restoration of chiral symmetry. The right panel of Fig. \ref{condensate} displays the chemical potential dependence of masses near the CEP for $eB = 20\ m_\pi^2$. A very soft sigma mode emerges as its mass approaches zero at the CEP.
 
\subsection{Spectral function at finite temperature}
We now turn to the meson spectral functions. For comparison, we first present the spectral function at 
$T=10$ MeV under zero density and magnetic field—conditions that closely mimic the vacuum. The result is consistent with that reported in \cite{Tripolt:2014aa}. However, since a different $p_3-$regulator is used in this work compared to \cite{Tripolt:2014aa}, it is necessary to first show the vacuum spectral function for reference. The pion spectral function displays a sharp peak near $\omega=100$ MeV, which corresponds to the pion pole mass. In the truncation scheme used here, this pole mass is smaller than the corresponding screening mass. Such a discrepancy between pole and screening masses has also been reported in \cite{Tripolt:2014aa}. Although adopting a different set of UV parameters, as in \cite{Wen:2023aa, Wang:2017vis}, could yield the correct pole mass, we retain the current parameter set because it produces the proper screening mass and thus ensures the correct threshold behavior in the spectral function. For energies above $\omega\geq 2m_\psi\approx 600$ MeV, an off-shell pion can decay into a quark–antiquark pair, which becomes kinematically allowed. This leads to a growth in the imaginary part of $\Gamma^{(2)}_\omega[\pi]$, and consequently enhances the pion spectral function for $\omega\geq 2m_\psi\approx 600$ MeV. On the other hand, the sigma spectral function shows a broad peak at $\omega=280$ MeV, which arises from the opening of the decay channel  $\sigma'\rightarrow \pi\pi$ near the threshold at $\omega=275$ MeV, with $\phi'$ indicating an off-shell meson and $\phi$ an on-shell meson. The imaginary part of $\Gamma^{(2)}_\omega[\sigma]$ increases significantly around this energy, coinciding with a zero-crossing in the real part of $\Gamma^{(2)}_\omega[\sigma]$.
\begin{figure}[H]\centering
\includegraphics[height=0.3\textwidth]{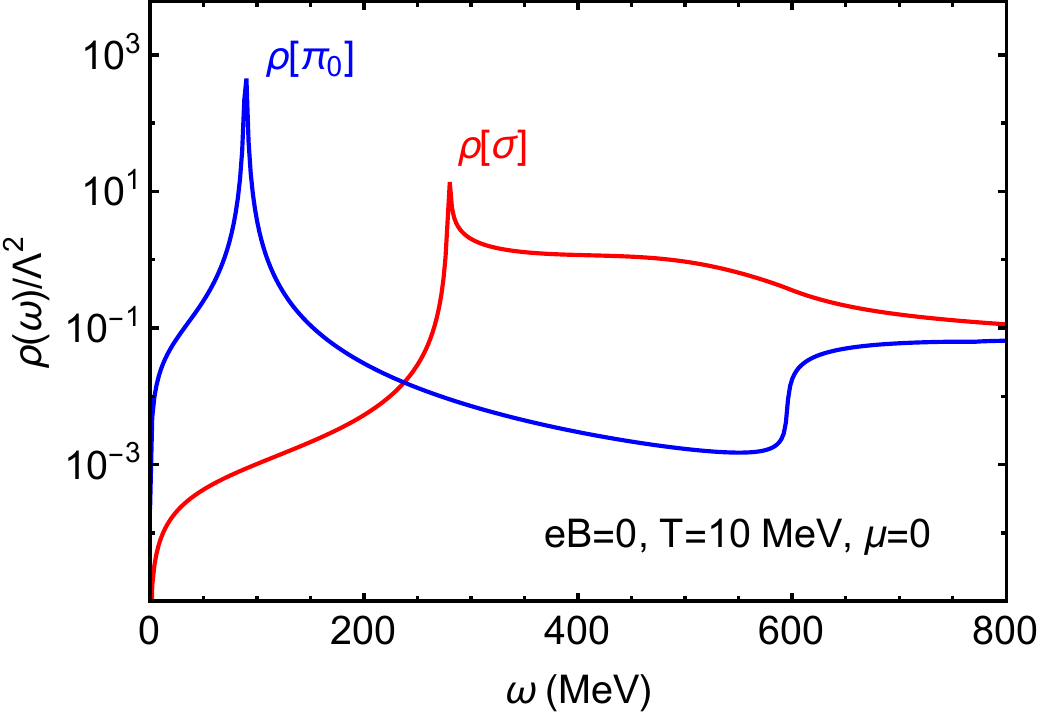}
\caption{Spectral function of sigma and pion at $T=10$ MeV and vanishing density and magnetic field. }
\label{spectral_0}
\end{figure}

\begin{figure}[H]
\centering
\includegraphics[height=0.3\textwidth]{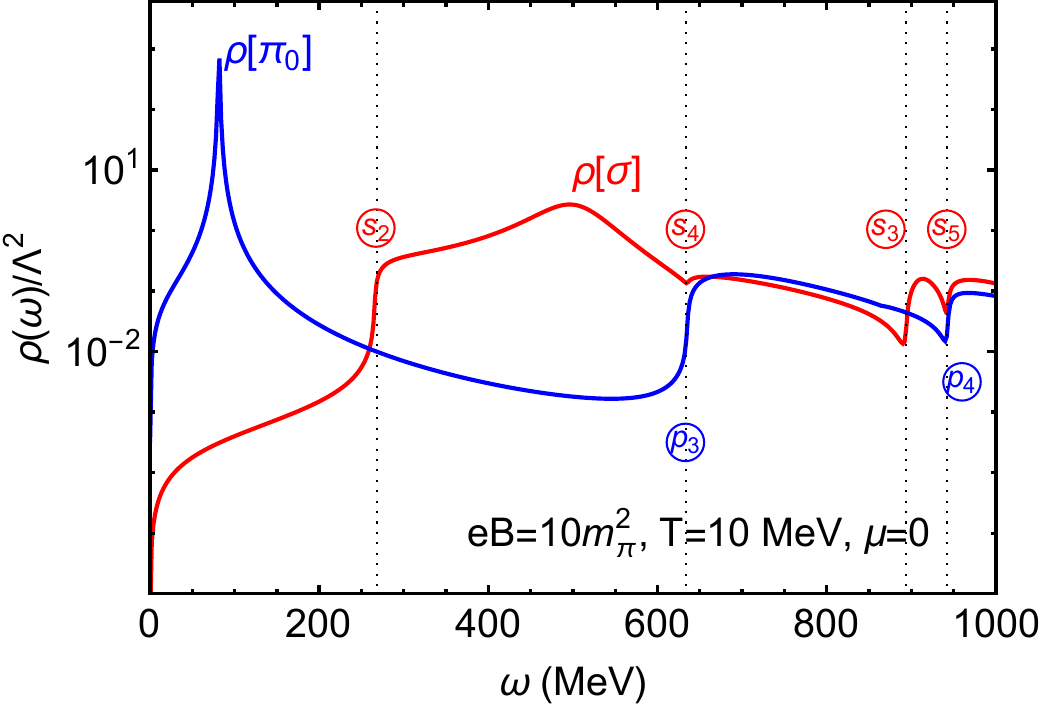}
\includegraphics[height=0.3\textwidth]{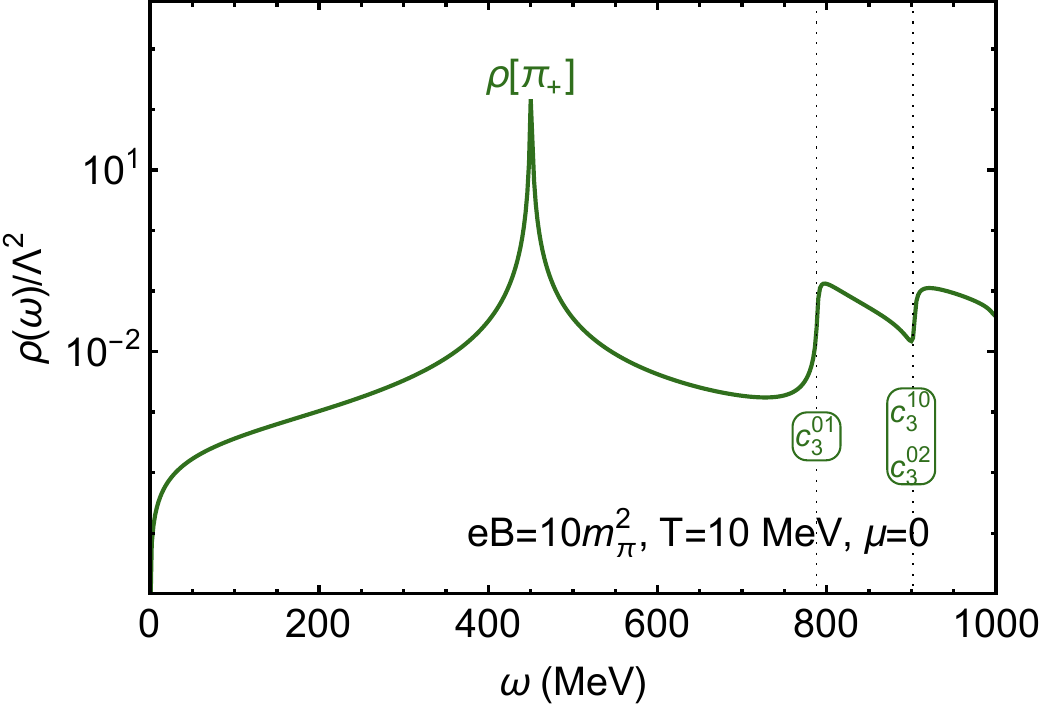}
\includegraphics[height=0.3\textwidth]{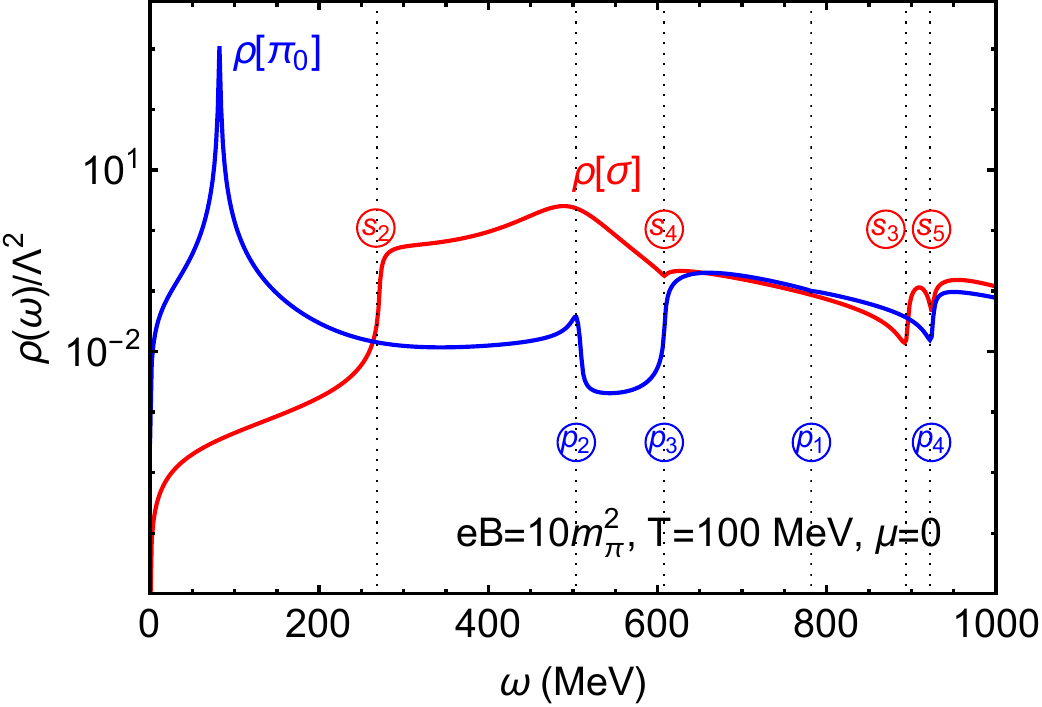}
\includegraphics[height=0.3\textwidth]{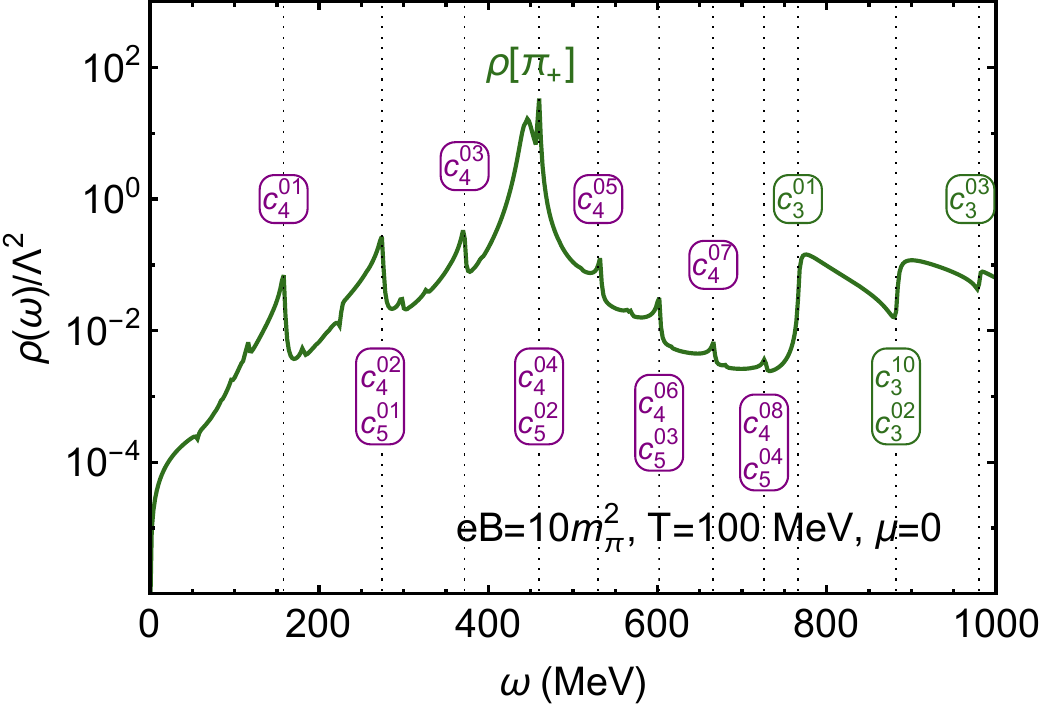}
\includegraphics[height=0.3\textwidth]{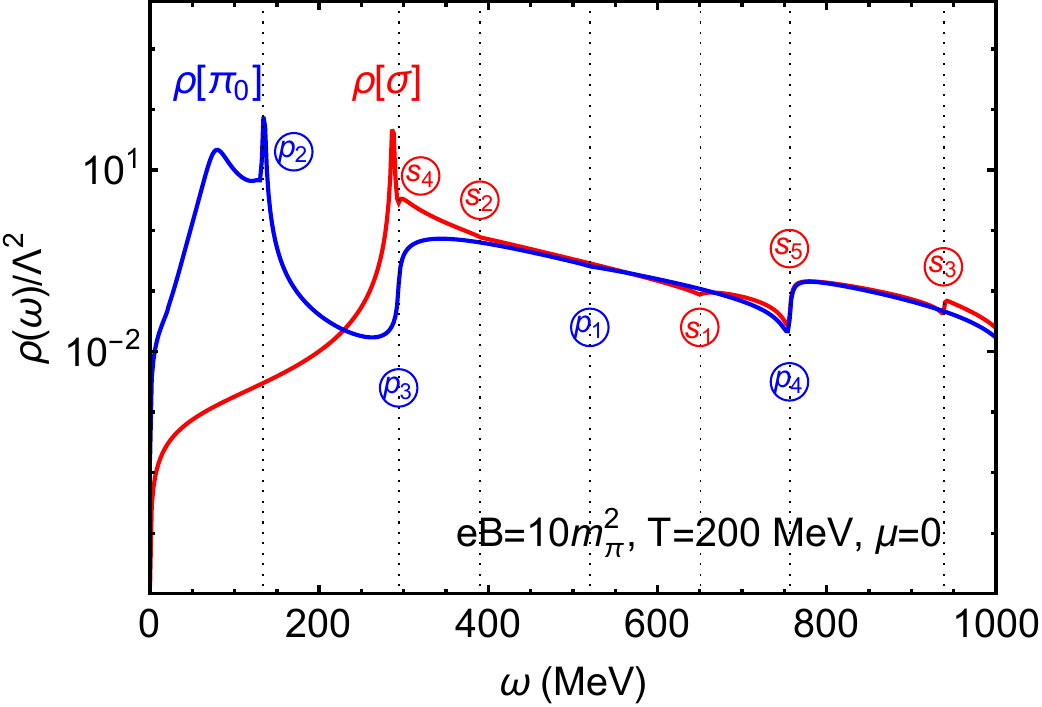}
\includegraphics[height=0.3\textwidth]{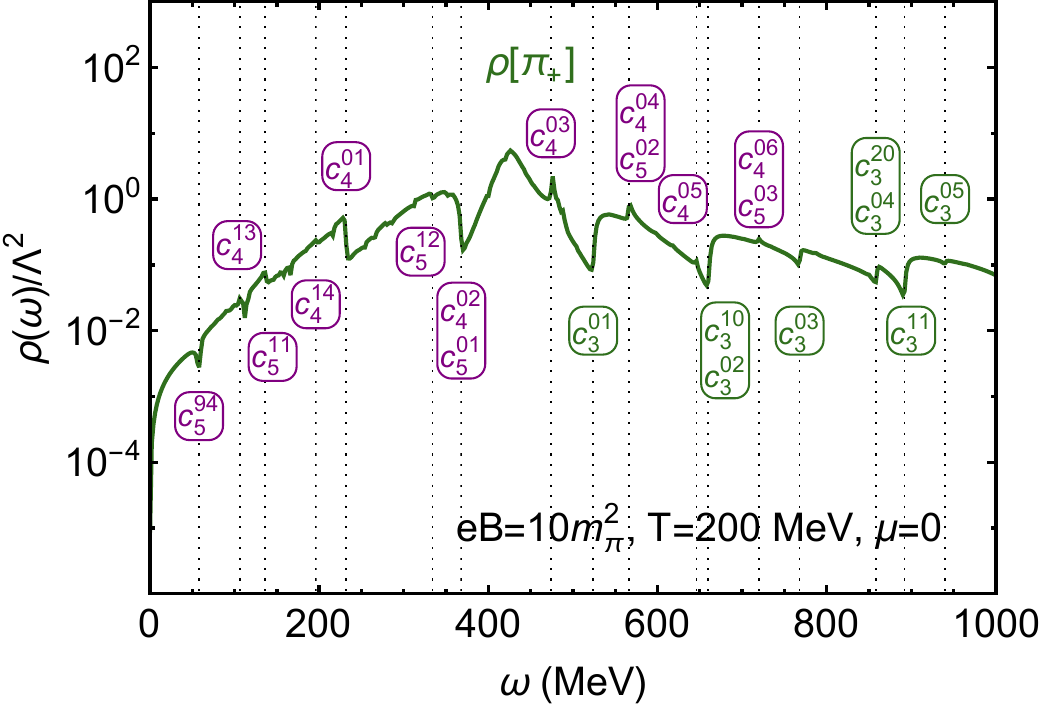}
\includegraphics[height=0.3\textwidth]{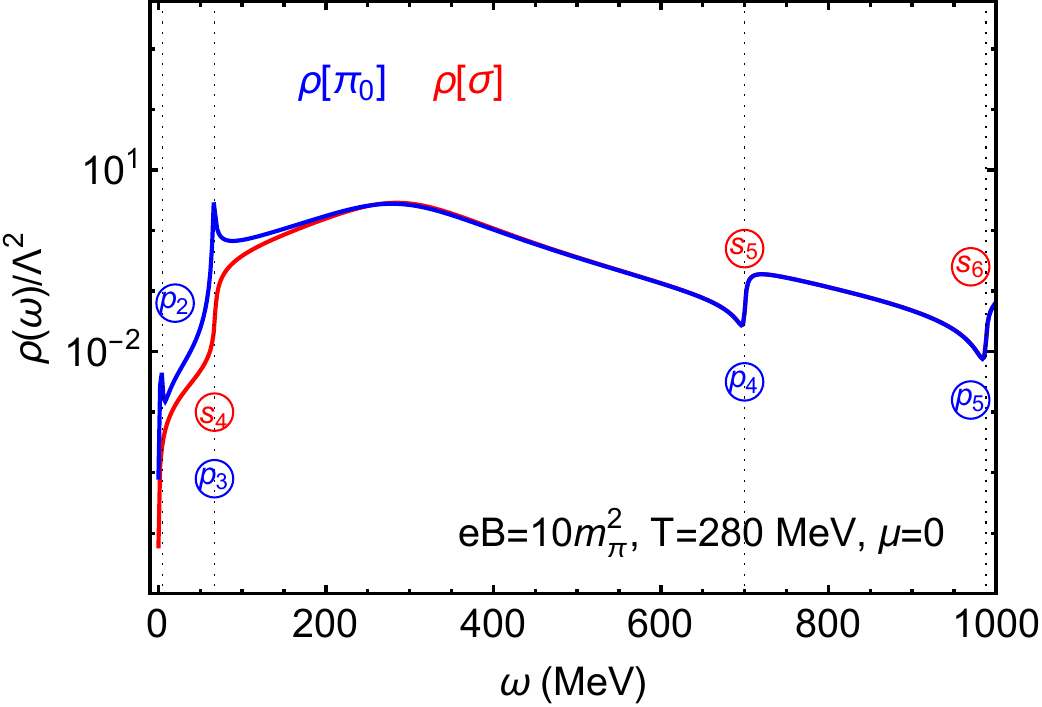}
\includegraphics[height=0.3\textwidth]{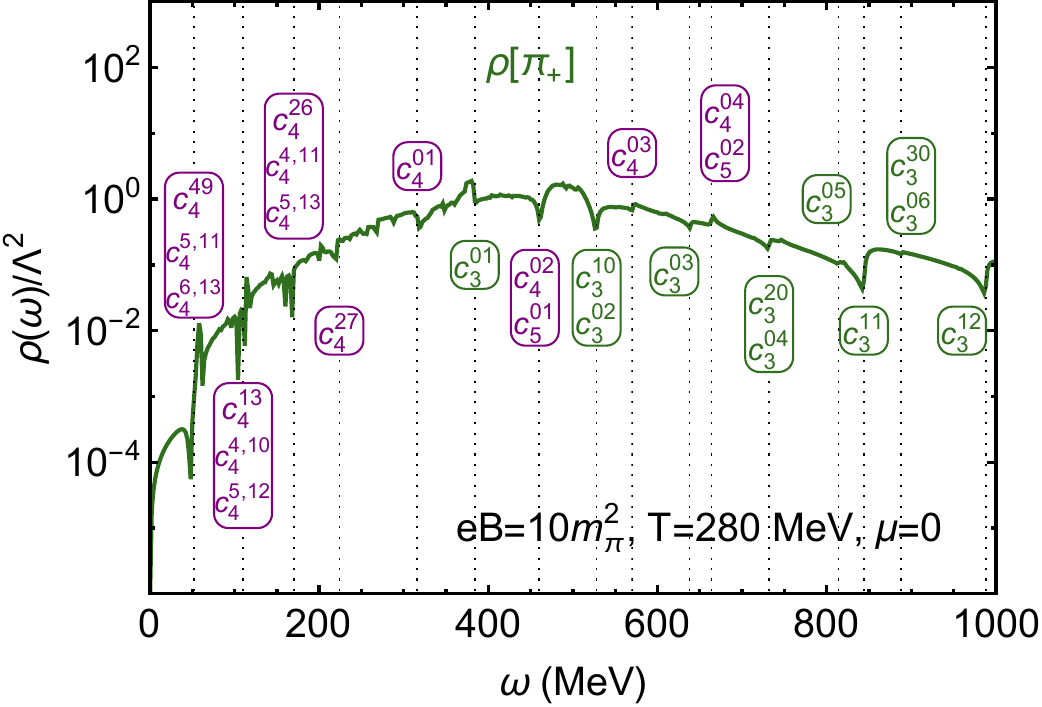}
\caption{Spectral functions of $\sigma$, $\pi_0$ (left panels) and $\pi_+$ (right panels) are shown versus external energy $\omega$ at $\mu=0$ and magnetic field $eB=10m_\pi^2$ for different temperature. Inserted annotations refer to the different processes affecting the spectral functions at the so indicated values of $\omega$, $s_1:\sigma'\rightarrow \sigma\sigma$, $s_2:\sigma'\rightarrow \pi_0\pi_0$, $s_3:\sigma'\rightarrow \pi_+\pi_-$, $s_4:\sigma'\rightarrow \psi_0\bar{\psi}_0$, $s_5:\sigma'\rightarrow d_1\bar{d}_1$, $s_6:\sigma'\rightarrow u_1\bar{u}_1$; $p_1:\pi_0'\rightarrow \pi_0\sigma$, $p_2:\pi_0' \pi_0\rightarrow\sigma$, $p_3:\pi_0'\rightarrow \psi_0\bar{\psi}_0$, $p_4:\pi_0'\rightarrow d_1\bar{d}_1$, $p_5:\pi_0'\rightarrow u_1\bar{u}_1$; $c_1: \pi_+'\sigma\rightarrow\pi_+$, $c_2: \pi_+'\pi_-\rightarrow\sigma$, $c_3^{ij}:\pi_+'\rightarrow u_i \bar{d}_j$, $c_4^{ij}:\pi_+'\bar{u}_i \rightarrow \bar{d}_j$, $c_5^{ij}:\pi_+'{d}_i\rightarrow u_j $.}
\label{spectral_T}
\end{figure}

After turning on the magnetic field, $\pi_0$ and $\pi_\pm$ are no longer degenerate. We show in Fig.\ref{spectral_T} the spectral functions of $\sigma$, $\pi_0$, and $\pi_+$ at zero density under a magnetic field $eB = 10~m_\pi^2$ for various temperature. The in top panel, at $T = 10$ MeV, the $\pi_0$ spectral function (blue curve) has a sharp peak at $\omega = 82\text{ MeV}$, originating from the zero-crossing of the real part of $\Gamma^{(2)}_\omega[\pi_0]$. Compared to the $B = 0$ case in Fig.\ref{spectral_0}, this peak shifts to a lower $\omega$ value, consistent with the decrease in the $\pi_0$ curvature mass as the magnetic field increases. The spectral function of the $\sigma$ meson at $T = 10$ MeV and $eB = 10~m_\pi^2$ is shown as the red curve in the top panel. The peak shifts to a higher energy of $\omega \sim 504\text{ MeV}$ compared to the $B = 0$ case in Fig. \ref{spectral_0}, which is consistent with the increase in the $\sigma$ screening mass under stronger magnetic fields. In addition, the peak becomes broader relative to the $B = 0$ case. This broadening occurs because the decay channel $\sigma' \rightarrow \pi_0\pi_0$ opens around $\omega = 268\text{ MeV}$, leading to a rise in the imaginary part of $\Gamma^{(2)}_\omega[\sigma]$ before the real part undergoes zero-crossing. An enhancement emerges in the $\pi_0$ and $\sigma$ spectral function around $\omega = 633\text{ MeV}$, due to the opening of the decay channel $\pi_0' \rightarrow \psi_0\bar{\psi}_0$ and $\sigma' \rightarrow \psi_0\bar{\psi}_0$, where $\psi_0$ denotes a u or d quark in the lowest Landau level and $\bar{\psi}_0$ the corresponding anti-quark. A similar increase occurs around $\omega = 960\text{ MeV}$, resulting from the decay $\pi_0' \rightarrow d_1\bar{d}_1$ and $\sigma' \rightarrow d_1\bar{d}_1$, which corresponds to an off-shell $\pi_0$ and $\sigma$ decaying into a d-quark and an anti-d-quark pair in the first Landau level. It is important to note that, as indicated by the flow equations in \eqref{twopoint_neutral}, both $\pi_0$ and $\sigma$ can only decay into quark-antiquark pairs occupying the same Landau level. Decay channels to mesons also contribute at high energy, which includes $\pi_0' \rightarrow \pi_0 \sigma$ around $\omega = 864\text{ MeV}$, where an off-shell $\pi_0$ decays into an on-shell $\pi_0$ and a $\sigma$.  Additionally, the decay channel $\sigma' \rightarrow \pi_+\pi_-$ contributes to the $\sigma$ spectral function around $\omega = 894\text{ MeV}$.

In the spectral function of $\pi_+$ at $T=10$ MeV under a magnetic field of $eB = 10m_\pi^2$, represented by the green curve in top panel of Fig. \ref{spectral_T}, a sharp peak is observed at $\omega \sim 450\text{ MeV}$, corresponding to the mass of $\pi_+$, which is consistent with $\sqrt{(m_{\pi_0}^{\text{pole}})^2 + eB}$. The spectral function is enhanced when decay into quark–antiquark pairs becomes kinematically allowed, specifically through channels of the form $\pi_+' \rightarrow u_i\bar{d}_j$, where an off-shell charged pion decays into a u-quark in the $i$-th Landau level and a $\bar{d}$-quark in the $j$-th Landau level. According to the flow equation, a charged pion cannot decay into a quark and an antiquark both in the ground Landau level. However, decay into a pair both occupying the first Landau level is allowed; this occurs at $\omega = 1060\text{ MeV}$, though it lies outside the range shown in the figure. The dominant contribution at large energies comes from the channel $\pi_+' \rightarrow u_0\bar{d}_1$, in which the pion decays into a u-quark in the ground state and a $\bar{d}$-quark in the first Landau level. This occurs at $\omega = 788\text{ MeV}$. Similarly, at $\omega = 902\text{ MeV}$, both $\pi_+' \rightarrow u_1\bar{d}_0$ and $\pi_+' \rightarrow u_0\bar{d}_2$ can occur, resulting in further enhancement of the spectral function.

At $T = 100\text{ MeV}$, additional interesting structures emerge in the spectral functions due to thermal effects. The second row of Fig. \ref{spectral_T} shows the spectral functions of the neutral mesons $\pi_0$ (blue curve) and $\sigma$ (red curve). At this temperature, chiral symmetry remains broken, and the overall shapes of the $\pi_0$ and $\sigma$ spectral functions are similar to those at $T = 10\text{ MeV}$. A new feature appears in the $\pi_0$ spectral function at $\omega = 511\text{ MeV}$, which is attributed to the annihilation process $\pi_0' \pi_0 \rightarrow \sigma$, where an off-shell $\pi_0$ interacts with a thermally excited on-shell $\pi_0$ to produce an on-shell $\sigma$. As this process relies on thermal excitation, it occurs only at finite temperatures.

The $\pi_+$ spectral function at $T = 100\text{ MeV}$, presented as the green curve in the second row of Fig. \ref{spectral_T}, retains a general shape similar to that at $T = 10\text{ MeV}$, but exhibits a multi-peak structure. The main peak at $\omega = 446\text{ MeV}$, resulting from the zero-crossing of the real part of $\Gamma^{(2)}_\omega[\pi_+]$, corresponds to the pole mass of $\pi_+$. As in the $T = 10\text{ MeV}$ case, a series of enhancements occur at higher energies ($\omega > 766\text{ MeV}$) due to decays into quark-antiquark pairs across different Landau levels, i.e., $\pi_+' \rightarrow u_i \bar{d}_j$. At lower energies, before these decay channels open, a multi-peak structure arises from processes involving annihilation with thermally excited quarks: specifically, $\pi_+' \bar{u}_i \rightarrow \bar{d}_j$ and $\pi_+' d_i \rightarrow u_j$. These processes occur only at finite temperature and become more significant as temperature increases. They induce minor changes in the real and imaginary parts of the spectral function, producing small peaks, as illustrated in Fig.\ref{real_imag_part} in Appendix.\ref{App1}. Since these modifications are modest and do not alter the dominant structure, the corresponding peaks are relatively low in amplitude. For example, the peak annotated as $c_4^{01}$ at $\omega = 158\text{ MeV}$ corresponds to $\pi_+' \bar{u}_0 \rightarrow \bar{d}_1$, where an off-shell $\pi_+'$ annihilates with a thermally excited $\bar{u}$-quark in the lowest Landau level to produce a $\bar{d}$-quark in the first Landau level. Similarly, $c_4^{02}$ and $c_5^{01}$ at the same energy denote the processes $\pi_+' \bar{u}_0 \rightarrow \bar{d}_2$ and $\pi_+' d_0 \rightarrow u_1$, respectively, which share the same threshold energy. A sequence of such processes occurs at various energies, as annotated in Fig. \ref{spectral_T}. Notably, all these involve a quark or antiquark in the lowest Landau level, since such excitations dominate the thermal bath. Processes involving higher Landau level quarks are suppressed and contribute minimally to the spectral function—though they may become more relevant at even higher temperatures.

At a temperature of $T=200$MeV and under a magnetic field of $eB = 10m_\pi^2$, as shown in the third row of Fig. \ref{spectral_T}, the $\sigma$ spectral function (red curve) exhibits a peak at $\omega=286$MeV corresponding to the zero-crossing of the real part of $\Gamma^{(2)}_\omega[\sigma]$. Compared to the lower-temperature case, this sigma peak is sharper and narrower, indicating a more stable sigma meson near the crossover temperature. At this energy, decay into two pions or a quark–antiquark pair is not allowed. Just above the peak energy, decay into a ground-state quark–antiquark pair becomes possible, leading to a broadening of the peak. As the energy increases further, additional decay channels open, such as those into two $\sigma$ mesons, two $\pi_0$ mesons, a pair of $\pi_\pm$ mesons, and quark–antiquark pairs in the first Landau level, resulting in further enhancement of the spectral function. The $\pi_0$ spectral function (blue curve) shows a peak near $\omega=82$ MeV due to the zero-crossing of the real part of $\Gamma^{(2)}_\omega [\pi_0]$, followed by another peak around  $\omega=134$ MeV. The latter originates from the process where an off-shell $\pi_0'$ annihilates with a thermally excited $\pi_0$ from the thermal medium to produce a $\sigma$ meson. At higher energies, above the opening of the $\pi_0'\rightarrow \psi_0\psi_0$ decay channel near $\omega=300$ MeV, the structure of the $\pi_0$ spectral function becomes qualitatively similar to that of the $\sigma$ meson.

The $\pi_+$ spectral function at $T=200$MeV, represented by the green curve in the third row of Fig. \ref{spectral_T}, is overall lower and broader compared to the corresponding spectral function at $T=100$ MeV. It is also broader than the $\sigma$ and $\pi_0$ spectral functions at $T=200$ MeV, suggesting that $\pi_+$ mesons are more readily dissolved in the thermal medium under a magnetic field. This behavior is associated with the various decay and annihilation channels available to the $\pi_+$, as annotated in the  boxes in the right panel. The main peak around $420$ MeV arises from the zero-crossing of the real part of $\Gamma^{(2)}_\omega [\pi_+]$. Above $\omega=524$ MeV, decays into $u_i\bar{d}_j$ quark pairs become allowed, each with a distinct threshold energy labeled as $c_{3}^{ij}$ in the figure. At lower energies, the structure of the $\pi_+$ spectral function is influenced by annihilation processes between a $\pi_+'$ and thermally excited quarks or antiquarks. These annihilation channels contribute only weakly to the real and imaginary parts of $\Gamma^{(2)}_\omega [\pi_+]$, making the corresponding peaks less prominent or invisible once decay channels open. Since thermally excited quarks in high Landau levels are suppressed, the dominant contributions to annihilation come from quarks and antiquarks in the lowest Landau level. Thus, between $\omega\approx232$ MeV and $\omega\approx720$ MeV, all relevant annihilation processes involve at least one quark or antiquark in the lowest Landau level. At even lower energies, where only annihilation channels associated with higher Landau levels are accessible, the spectral function displays an oscillatory-like structure. This arises from the dense spacing of threshold energies for annihilation processes involving quarks in higher Landau levels. A few such channels related to the first and higher Landau levels are annotated at low energies; many more contribute in this region.
   
At an even higher temperature of $T=280$ MeV and for $eB = 10~m_\pi^2$, as shown in the last row of Fig. \ref{spectral_T}, the $\sigma$ and $\pi_0$ spectral functions become nearly degenerate, except in the very soft region. The $\pi_0$ spectral function exhibits two additional peaks at $\omega=4$MeV and $\omega=67$MeV, originating from the processes $\pi_0'\pi_0\rightarrow\sigma$ and $\pi_0'\rightarrow \psi_0\psi_0$, respectively. For both $\sigma$ and $\pi_0$, the main peak around $\omega=300$MeV is very broad, suggesting that the mesons exist only as loosely bound states. At higher energies, decay channels into $\bar{d}d$ and $\bar{u}u$ pairs in the first Landau level open around $700$MeV and $1$GeV, respectively, leading to an enhancement in the spectral function. The $\pi_+$ spectral function exhibits significant broadening at $T=280$ MeV and $eB = 10~m_\pi^2$ as shown in the last row of Fig. \ref{spectral_T}. This broadening causes its main peak around $490$ MeV to merge with the small peaks from other processes and become nearly indistinguishable. Due to the reduced fermion mass and the higher thermal energy, a greater number of decay and annihilation channels—particularly those involving higher Landau levels—are present compared to the $T=200$MeV case. However, their individual contributions are minor. These oscillatory structures in $\pi_+$ spectral function become more pronounced at high temperature, where more Landau levels are thermally populated. We have explicitly verified that they originate from physical thresholds associated with Landau-level-resolved processes, rather than numerical artifacts (see Appendix E for a detailed analysis). The oscillatory pattern can be understood as a sequence of threshold openings associated with Landau-level-resolved scattering and decay channels, whose density increases with temperature and cutoff scale.

 \begin{figure}[H]\centering
\includegraphics[height=0.25\textwidth]{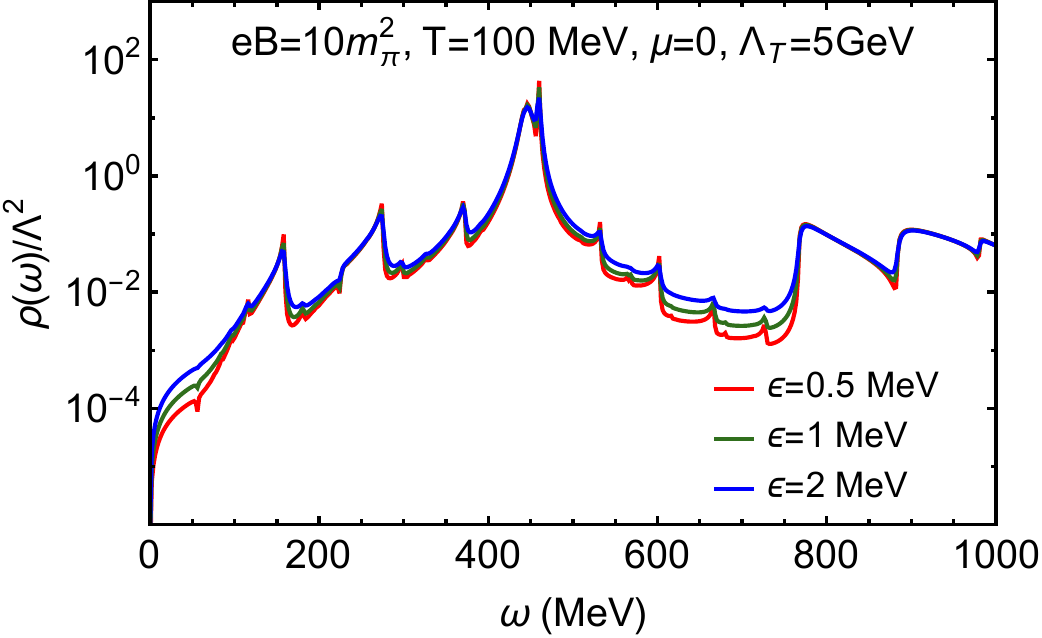}
\caption{Spectral function of $\pi_+$ at $T=100$MeV, $\mu=0$ and magnetic field $eB=10m_\pi^2$ with different values of $\epsilon=0.5$ MeV (red), $\epsilon=1$ MeV (green), $\epsilon=2$ MeV (blue).}
\label{app_T100_epsilon}
\end{figure}
In the end of this section, we briefly discuss the dependence on the regulator parameter $\epsilon$. In the analytic continuation from Euclidean to Minkowski space, a small imaginary part $\epsilon$ is introduced as a regulator. To assess its influence, we have computed the $\pi^+$ spectral function for different values of $\epsilon$ at fixed temperature. As shown in Fig.~\ref{app_T100_epsilon}, varying $\epsilon$ does not change the position of the spectral peaks, indicating that the pole structure is unaffected. However, increasing $\epsilon$ enhances the imaginary part of the two-point function, which smooths out fine structures in the spectral function and makes smaller peaks less distinguishable. In the present work, we fix $\epsilon = 1~\text{MeV}$, which provides a balance between numerical stability and resolution. With $\epsilon$ fixed, the broadening of the spectral function observed at higher temperature is therefore of physical origin, reflecting the increased number of available scattering and decay channels.

\subsection{Spectral function around the CEP}
We further examine how meson spectral functions depend on the chemical potential in a magnetic field. Our analysis focuses on a temperature of $T=56$ MeV and $eB=20m_\pi^2$, corresponding to a horizontal line in the phase diagram across the critical endpoint (CEP). This choice allows us to study behavior near the CEP while maintaining significant thermal effects. We did not choose $eB=10m_\pi^2$ because its CEP occurs at a very low temperature ($T=25$ MeV), where thermal excitations are negligible. Nor did we choose $eB=30m_\pi^2$, as the thresholds for key annihilation and decay channels would lie beyond our energy range of interest.

\begin{figure}[H]
\centering
\includegraphics[height=0.3\textwidth]{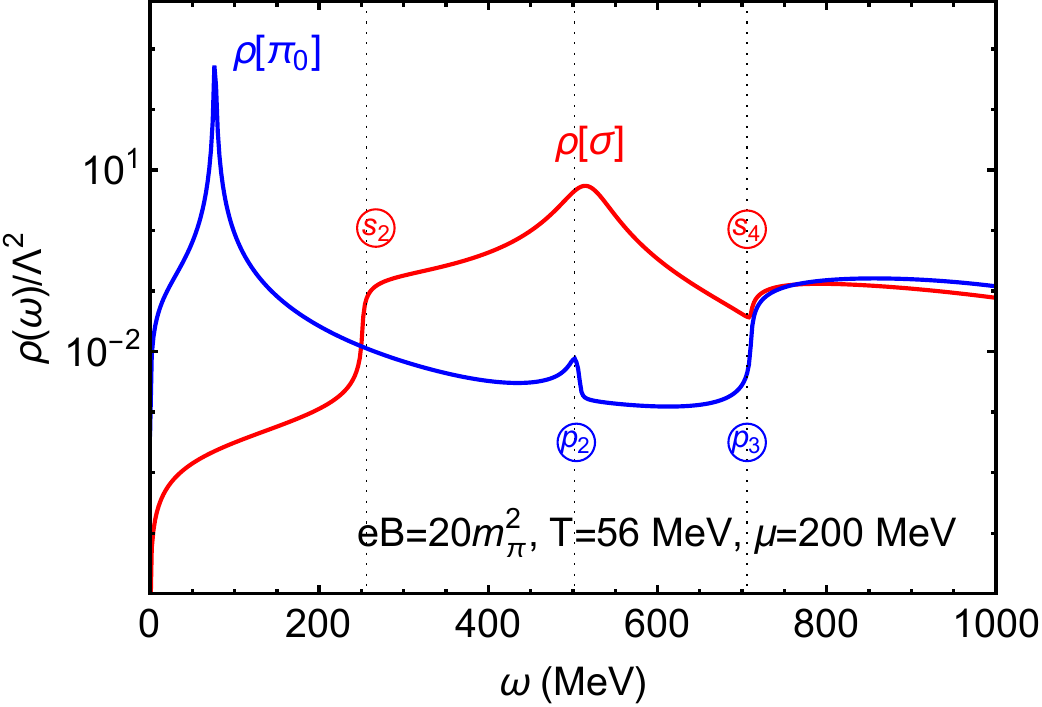}
\includegraphics[height=0.3\textwidth]{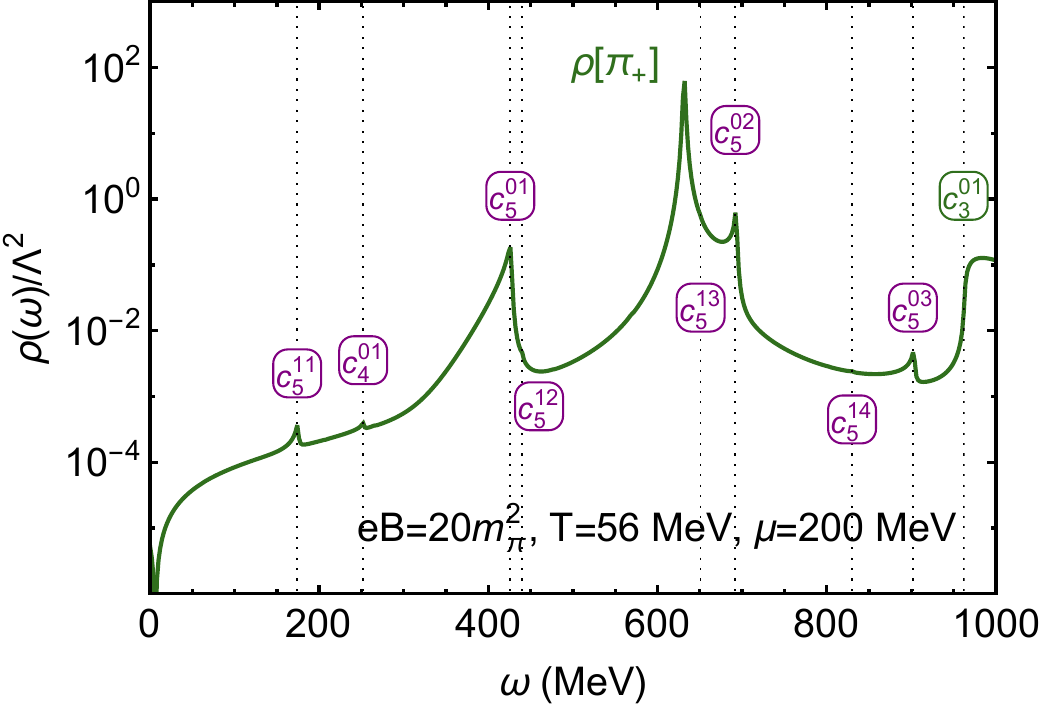}
\includegraphics[height=0.3\textwidth]{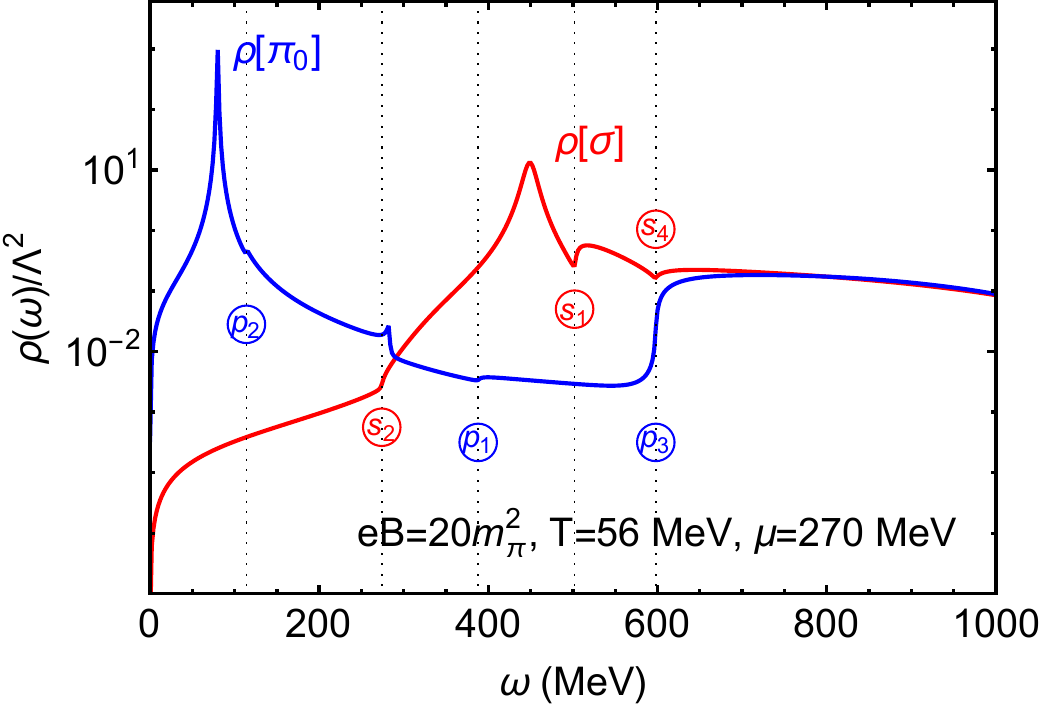}
\includegraphics[height=0.3\textwidth]{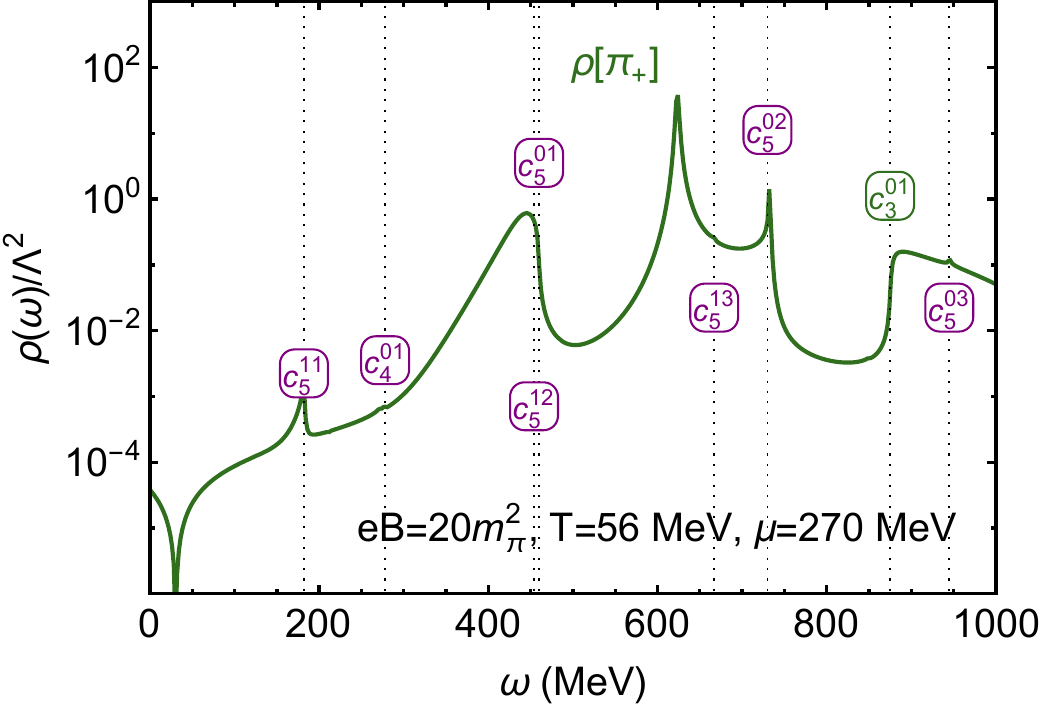}
\includegraphics[height=0.3\textwidth]{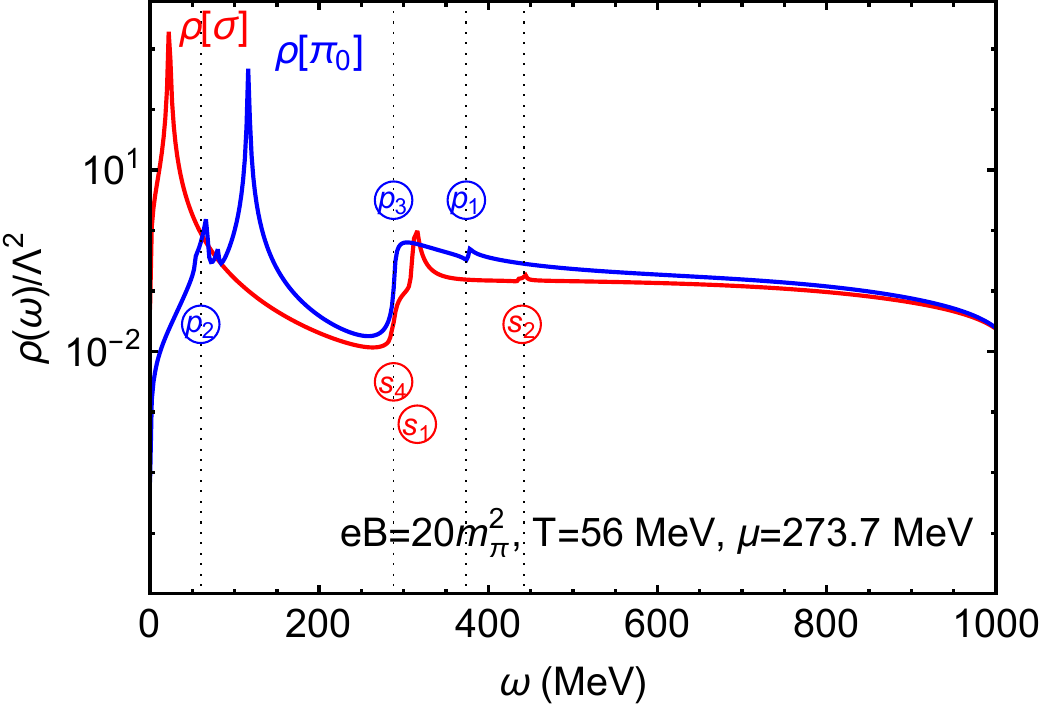}
\includegraphics[height=0.3\textwidth]{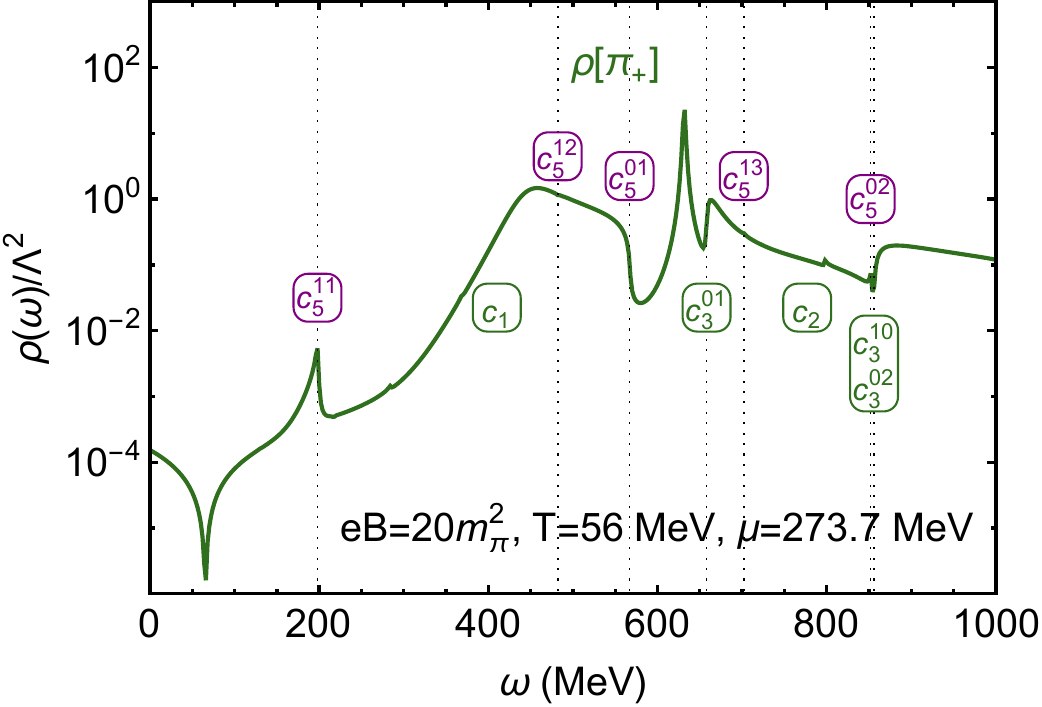}
\includegraphics[height=0.3\textwidth]{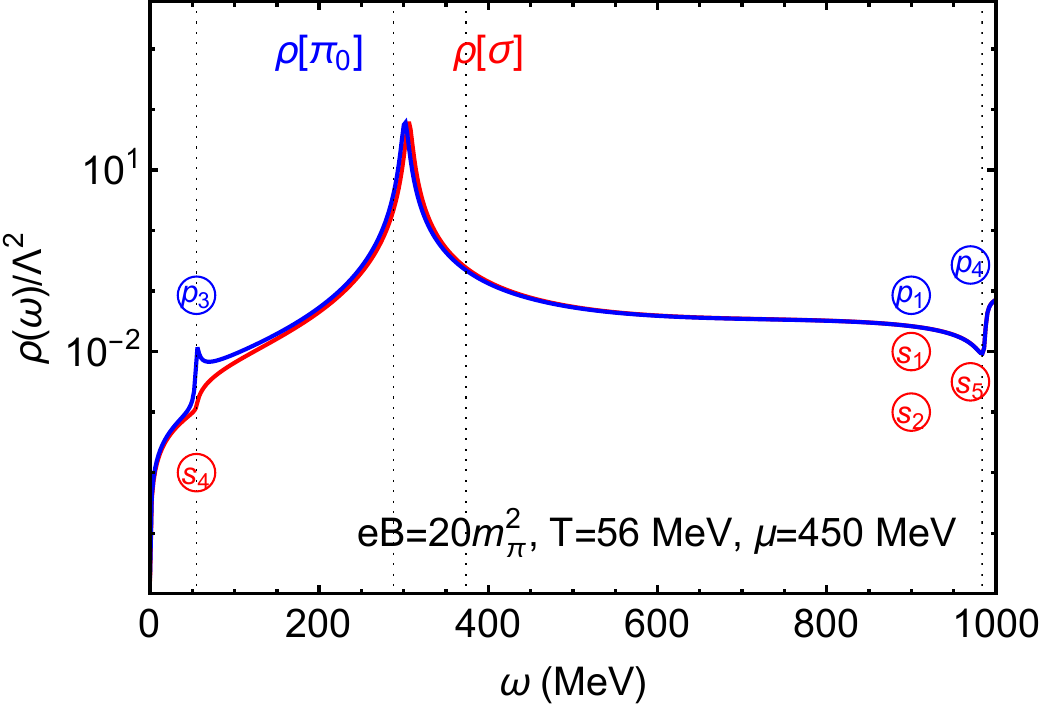}
\includegraphics[height=0.3\textwidth]{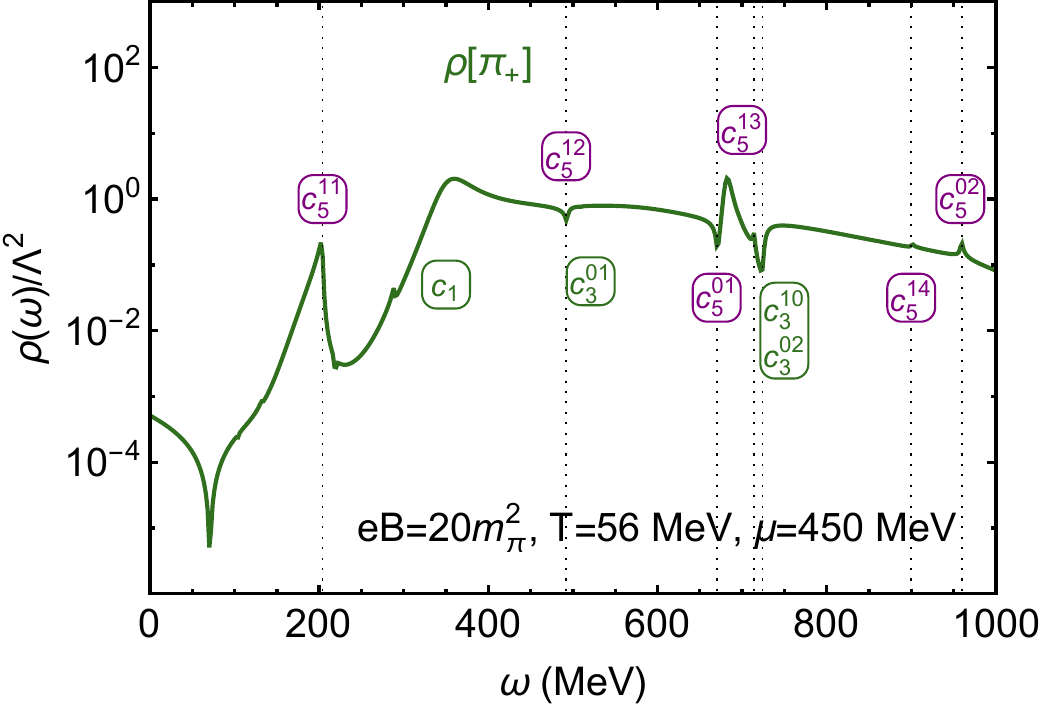}
\caption{Spectral functions of $\sigma$, $\pi_0$ (left panel) and $\pi_+$ (right panel) are shown versus external energy $\omega$ at $T=56$MeV and magnetic field $eB=20m_\pi^2$ for various chemical potential $\mu$.  Inserted annotations refer to the different processes affecting the spectral functions at the so indicated values of $\omega$, $s_1:\sigma'\rightarrow \sigma\sigma$, $s_2:\sigma'\rightarrow \pi_0\pi_0$, $s_3:\sigma'\rightarrow \pi_+\pi_-$, $s_4:\sigma'\rightarrow \psi_0\bar{\psi}_0$, $s_5:\sigma'\rightarrow d_1\bar{d}_1$, $s_6:\sigma'\rightarrow u_1\bar{u}_1$; $p_1:\pi_0'\rightarrow \pi_0\sigma$, $p_2:\pi_0' \pi_0\rightarrow\sigma$, $p_3:\pi_0'\rightarrow \psi_0\bar{\psi}_0$, $p_4:\pi_0'\rightarrow d_1\bar{d}_1$, $p_5:\pi_0'\rightarrow u_1\bar{u}_1$; $c_1: \pi_+'\sigma\rightarrow\pi_+$, $c_2: \pi_+'\pi_-\rightarrow\sigma$, $c_3^{ij}:\pi_+'\rightarrow u_i \bar{d}_j$, $c_4^{ij}:\pi_+'\bar{u}_i \rightarrow \bar{d}_j$, $c_5^{ij}:\pi_+'{d}_i\rightarrow u_j $.}
\label{spectral_mu}
\end{figure}

Over a wide range of chemical potentials, the spectral functions remain largely unchanged, which is consistent with the Silver Blaze property \cite{Cohen:2003kd}. At $\mu=200$ MeV, the neutral meson spectral functions in the top left panel of Fig.\ref{spectral_mu} resemble those at  $\mu=0$ and $T=100$MeV in Fig. \ref{spectral_T}, except that the decay channel into quark-antiquark pairs in the first Landau level is absent here. Due to the stronger magnetic field, the threshold for this channel exceeds 1 GeV and thus lies outside the energy range shown. As the critical endpoint is approached, however, significant changes emerge, particularly in the sigma spectral function. At $\mu=270$ MeV, the sigma peak shifts to lower energy and narrows, signaling the formation of a more stable sigma meson near the CEP. Nevertheless, a discrepancy persists between the curvature mass and the pole mass: while the peak lies around $\omega=450$ MeV, the threshold for the decay $\sigma'\rightarrow\sigma\sigma$ has already dropped to about $500$ MeV. At the critical endpoint ($\mu=273.7$ MeV), the sigma meson becomes a stable soft mode with a sharp peak near $\omega=20$ MeV. In contrast, the $\sigma'\rightarrow\sigma\sigma$ threshold determined by the curvature mass—lags behind. This behavior stems from the extreme sensitivity of the pole mass to higher-order derivatives of the effective potential, combined with the rapid variation of the sigma curvature mass near the CEP. Upon further increasing the chemical potential, the sigma and pion spectral functions become degenerate, similar to the high-temperature case. A key difference, however, is that the spectral functions now display sharp peaks rather than broad resonances, indicating the emergence of stable dynamical quasi-particles.

The $\pi_+$ spectral function, shown in the right column of Fig. \ref{spectral_T}, also exhibits a multi-peak structure, with distinct features compared to the neutral mesons. While the neutral meson spectral functions remain nearly identical at $\mu=200$ MeV and $\mu=0$, the $\pi_+$ spectral function shows clear differences when compared to Fig. \ref{spectral_T56_B20}. At $\mu=200$ MeV, the multi-peak structure is dominated by annihilation processes of the type $\pi_+'{d}_i\rightarrow u_j$, labeled as $c_5^{ij}$. In contrast, contributions from $\pi_+'\bar{u}_i \rightarrow \bar{d}_j$ ($c_4^{ij}$) are negligible. This asymmetry arises because, at finite chemical potential, the medium is populated mainly by thermally excited quarks, while antiquarks are highly suppressed. As a result, with increasing chemical potential, processes like $\pi_+'{d}_i\rightarrow u_j$ are enhanced, leading to larger and more prominent peaks; while those involving antiquarks are further suppressed, becoming weaker or even invisible. This effect can be seen, for example, by comparing the magnitudes of the $c_4^{01}$ and $c_5^{01}$ peaks in the $\pi_+$ spectral function at $\mu=200$ MeV. Additionally, the peak labeled $c_5^{11}$ grows with increasing chemical potential, reflecting the enhanced probability of such processes in a high-density environment. As $\mu$ rises, the quark mass decreases, lowering the energy threshold for the decay channel $\pi_+'\rightarrow u_i \bar{d}_j$ and enabling contributions from higher Landau levels, as seen at $\mu=273.7$ MeV and $\mu=450$ MeV. Near the critical point, the sigma meson becomes soft, emerging as the lightest degree of freedom and the dominant thermal excitation in the medium. This gives rise to processes such as $\pi_+'\sigma\rightarrow\pi_+$, which enhance and broaden the spectral function around $\omega=400$MeV. Unlike the neutral mesons, whose spectral functions develop sharp peaks at high density, the charged pion is a broad resonance due to the multitude of enhanced annihilation and decay channels available in this regime.

\section{Summary}
\label{s4}
Although the mass spectrum of mesons in a magnetic field has been extensively studied, their dynamical properties in a magnetized medium remain relatively unexplored. In this work, we analyze the spectral functions of neutral and charged mesons at finite temperature and density under an external magnetic field, employing the two-flavor quark-meson model within the functional renormalization group (FRG) approach.

A key step in our analysis is the proper treatment of momentum relations at vertices for charged particles in the magnetic field. Based on this, we derive the flow equations for two-point functions of both neutral and charged mesons. The flow equation for the effective potential is solved using both the Taylor expansion around a fixed point and the grid method, with the two approaches yielding consistent results. Derivatives of the effective potential obtained during its flow evolution serve as inputs for integrating the flow equations of the two-point functions. Spectral functions are subsequently extracted via analytical continuation from Euclidean to Minkowski space.

These spectral functions offer crucial insights into the dynamical properties of mesons in a magnetized medium. They reveal how meson masses evolve during the chiral phase transition at finite temperature and density, with the $\sigma$ spectral function notably exhibiting a soft mode near the critical endpoint. The momentum relations at interaction vertices further elucidate the underlying mechanisms: while $\sigma$ and $\pi_0$ couple only to quark-antiquark pairs within the same Landau level, $\pi_+$ can couple to $u\bar{d}$ pairs across arbitrary Landau levels. This fundamental difference in coupling patterns leads to distinctly structured spectral functions for neutral and charged mesons under magnetic fields. Our results demonstrate that the spectral functions of $\sigma$ and $\pi_0$ mesons develop new features due to decay channels involving quark-antiquark pairs occupying different Landau levels. In contrast, the charged pion spectral function displays a characteristic multi-peak structure at finite temperatures, originating from the multiple annihilation and decay channels unique to $\pi_+$ in a magnetic environment. This multi-peak pattern becomes more pronounced in a finite-density medium, where the $\pi_+$ meson forms a broad resonance at lower temperatures and densities compared to neutral mesons. 

This multi-peak feature is expected to be universal for charged mesons under magnetic fields, suggesting the presence of numerous quasi-particle modes in the magnetized medium. This finding has significant implications for understanding transport properties in magnetized fluids.

\acknowledgments
We are grateful to Yangyang Tan, Shijun Mao, Shi Yin and Weijie Fu for fruitful discussions. The work is supported by NSFC grant Nos. 12005112 (Zy.W.).

\appendix
\section{Real and imaginary part of two-point function}
\label{App1}
\begin{figure}[H]\centering
\includegraphics[width=0.45\textwidth]{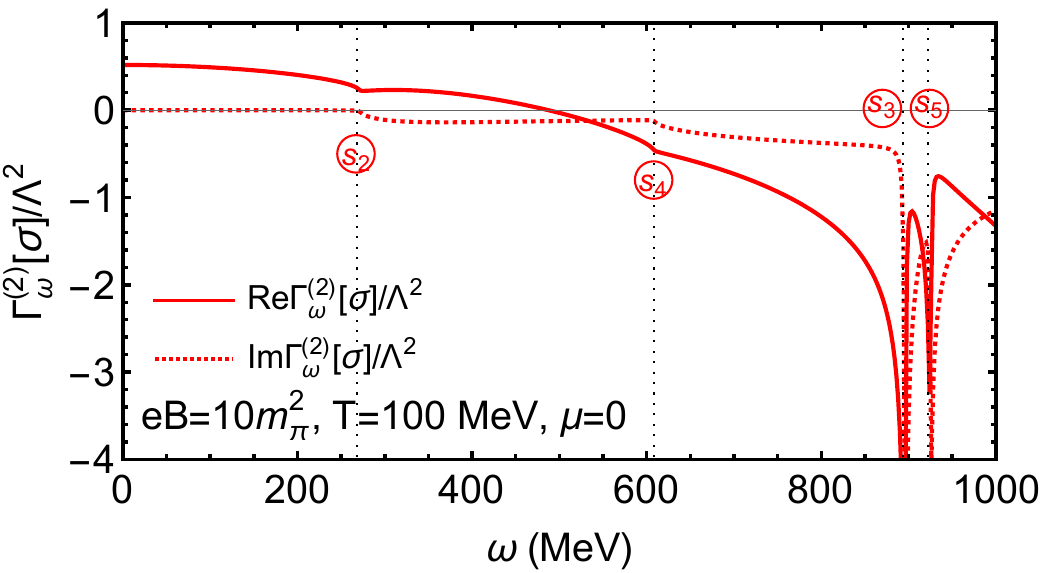}
\includegraphics[width=0.45\textwidth]{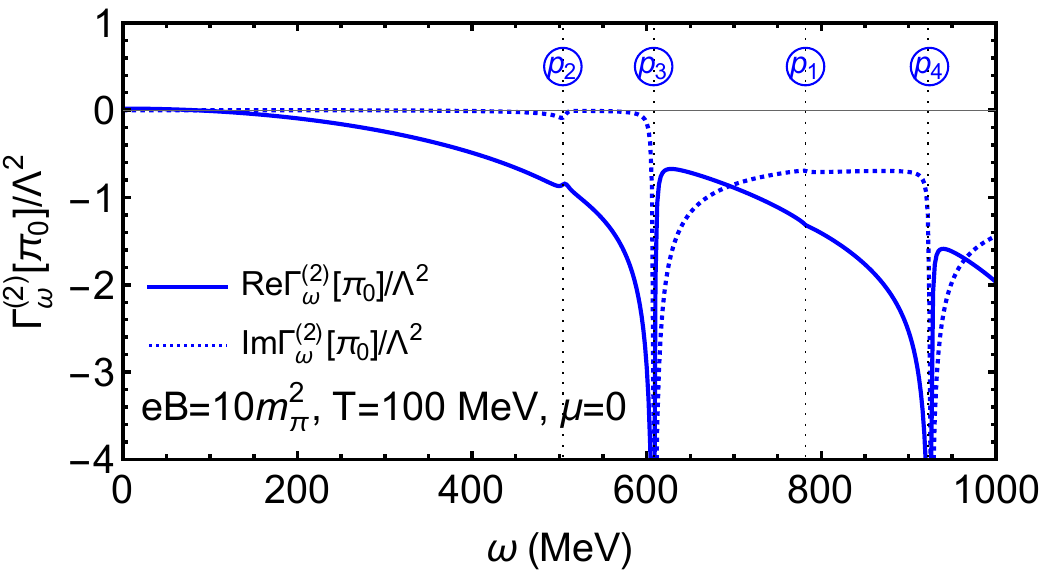}
\includegraphics[width=0.45\textwidth]{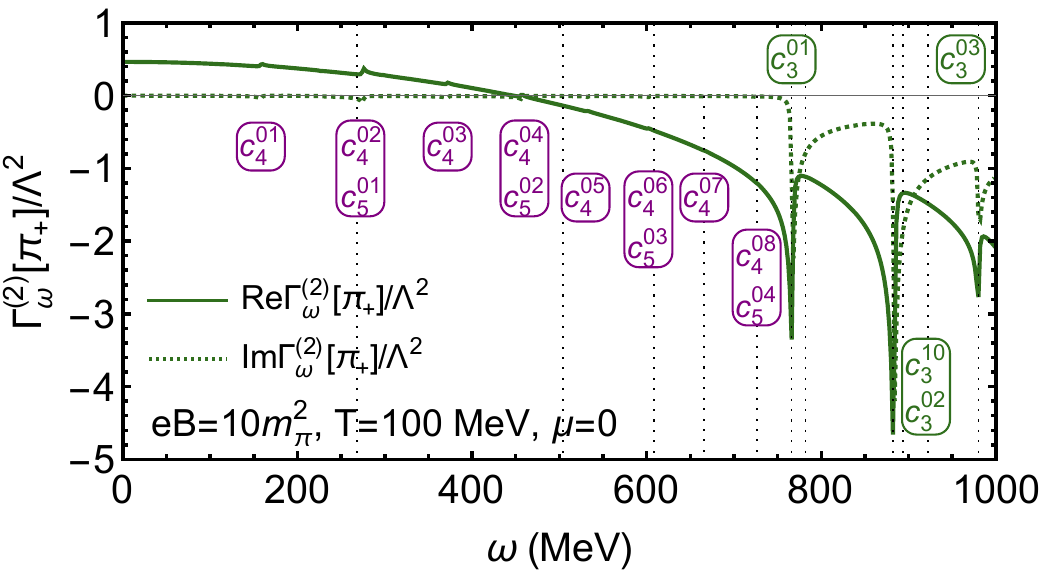}
\includegraphics[width=0.45\textwidth]{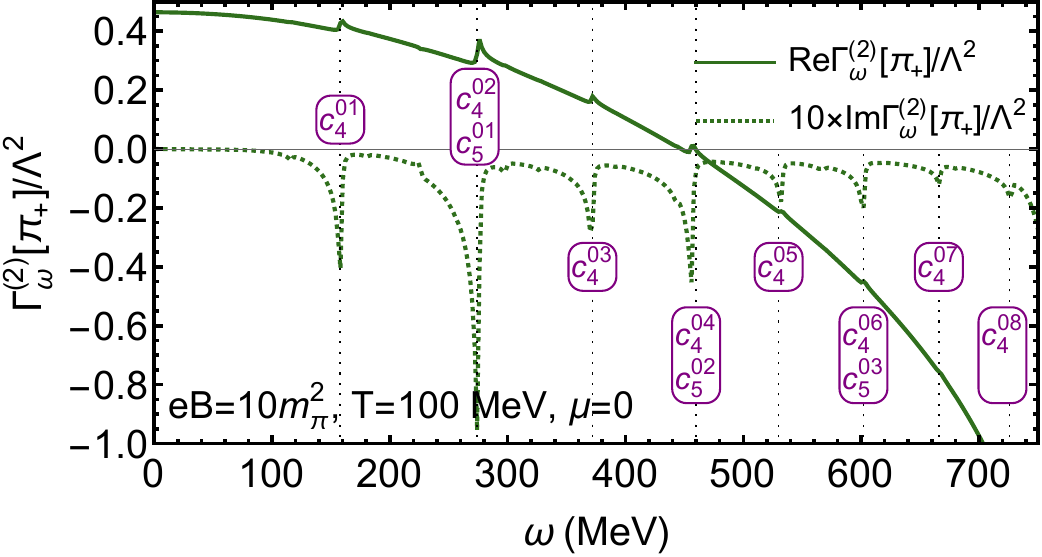}
\caption{Real part and imaginary part of $\sigma$, $\pi_0$ (upper two figures) and $\pi_+$ two-point functions (lower two figures) are shown versus external energy $\omega$ at $T=100$MeV, $\mu=0$ and magnetic field $eB=10m_\pi^2$. Inserted annotations refer to the different processes affecting the spectral functions at the so indicated values of $\omega$, $s_1:\sigma'\rightarrow \sigma\sigma$, $s_2:\sigma'\rightarrow \pi_0\pi_0$, $s_3:\sigma'\rightarrow \pi_+\pi_-$, $s_4:\sigma'\rightarrow \psi_0\bar{\psi}_0$, $s_5:\sigma'\rightarrow d_1\bar{d}_1$, $s_6:\sigma'\rightarrow u_1\bar{u}_1$; $p_1:\pi_0'\rightarrow \pi_0\sigma$, $p_2:\pi_0' \pi_0\rightarrow\sigma$, $p_3:\pi_0'\rightarrow \psi_0\bar{\psi}_0$, $p_4:\pi_0'\rightarrow d_1\bar{d}_1$, $p_5:\pi_0'\rightarrow u_1\bar{u}_1$; $c_1: \pi_+'\sigma\rightarrow\pi_+$, $c_2: \pi_+'\pi_-\rightarrow\sigma$, $c_3^{ij}:\pi_+'\rightarrow u_i \bar{d}_j$, $c_4^{ij}:\pi_+'\bar{u}_i \rightarrow \bar{d}_j$, $c_5^{ij}:\pi_+'{d}_i\rightarrow u_j $.}
\label{real_imag_part}
\end{figure}

To facilitate the interpretation of the structures observed in the spectral functions, we present the corresponding real and imaginary parts of the two-point functions. In Fig. \ref{real_imag_part}, we show the real and imaginary parts of the two-point functions for $\sigma$, $\pi_0$, and $\pi_+$ mesons at $T = 100\ \mathrm{MeV}$, $\mu = 0$, and $eB = 10\ m_\pi^2$, corresponding to the spectral functions displayed in Fig. \ref{spectral_T}.

The real part of the sigma 2-point function exhibits a zero-crossing at $\omega \approx 484\ \mathrm{MeV}$. However, this does not correspond to a stable sigma state, as the imaginary part becomes non-zero already for $\omega \geq 2m_{\pi_0} \approx 270\ \mathrm{MeV}$, due to the decay of a sigma meson into two neutral pions. At $\omega \geq 2m_{\psi} \approx 608\ \mathrm{MeV}$, the decay into a quark-antiquark pair in the lowest Landau level also contributes to the imaginary part. Furthermore, the decay channel $\sigma' \to \pi_+ \pi_-$, involving charged pions in the lowest Landau level, opens at $\omega \approx 894\ \mathrm{MeV}$. Since the sigma meson couples only to quark-antiquark pairs of the same flavor and same Landau level, the decay $\sigma' \to d_1 \bar{d}_1$ becomes possible at $\omega \approx 922\ \mathrm{MeV}$, where both the $d$-quark and $\bar{d}$-antiquark reside in the first Landau level.

The real part of the $\pi_0$ 2-point function shows a zero-crossing at $\omega \approx 90\ \mathrm{MeV}$, leading to a pronounced peak in the spectral function (Fig. \ref{spectral_T}), as the imaginary part remains negligible until $\omega \approx 504\ \mathrm{MeV}$, where the process $\pi_0' \pi_0 \to \sigma$ sets in. A small enhancement in both the real and imaginary parts appears for $\omega \geq m_\sigma + m_{\pi_0} \approx 782\ \mathrm{MeV}$, due to the decay into a sigma and a neutral pion. The decay into a quark-antiquark pair in the lowest Landau level contributes at $\omega \approx 608\ \mathrm{MeV}$, and the channel $\pi_0' \to d_1 \bar{d}_1$ opens at $\omega \approx 922\ \mathrm{MeV}$, analogous to the sigma case.

The lower two panels of Fig. \ref{real_imag_part} show the real and imaginary parts of the $\pi_+$ 2-point function. The real part displays a zero-crossing at $\omega \approx 446\ \mathrm{MeV}$, corresponding to a sharp peak in the spectral function (Fig. \ref{spectral_T}), since the imaginary part remains nearly zero until the first decay channel opens at $\omega \approx 766\ \mathrm{MeV}$. The dominant structures in the two-point function arise from the decay channels $\pi_+' \to u_i \bar{d}_j$, denoted by $c_3^{ij}$, where the $\pi_+$ meson decays into a $u$-quark in the $i$-th Landau level and a $\bar{d}$-antiquark in the $j$-th Landau level. Because the $\pi_+$ can couple to $u\bar{d}$ pairs across arbitrary Landau levels (except when both are in the lowest level), the first decay channel $u_0 \bar{d}_1$ opens at $766\ \mathrm{MeV}$, followed by $u_1 \bar{d}_0$ and $u_0 \bar{d}_2$ at $882\ \mathrm{MeV}$, and $u_0 \bar{d}_3$ at $980\ \mathrm{MeV}$. Since $u_1 \bar{d}_0$ and $u_0 \bar{d}_2$ contribute at the same threshold energy, their combined effect produces a change in the real and imaginary parts approximately twice as large as that from $u_0 \bar{d}_1$ at $766\ \mathrm{MeV}$.

To clarify the smaller variations below the decay threshold at $\omega \approx 766\ \mathrm{MeV}$, the imaginary part in the lower right panel is magnified by a factor of 10. The ability of $\pi_+$ to couple to $u\bar{d}$ pairs at arbitrary Landau levels enables a variety of annihilation processes, in which an off-shell $\pi_+'$ combines with a thermally excited $\bar{u}$ or $d$ quark to produce an on-shell $\bar{d}$ or $u$ quark in the magnetized thermal medium. These annihilation processes generate a sequence of small peaks in the real and imaginary parts of the two-point function, which in turn give rise to the multi-peak structure observed in the $\pi_+$ spectral function in Fig. \ref{spectral_T}.

 \begin{figure}[H]\centering
\includegraphics[width=0.45\textwidth]{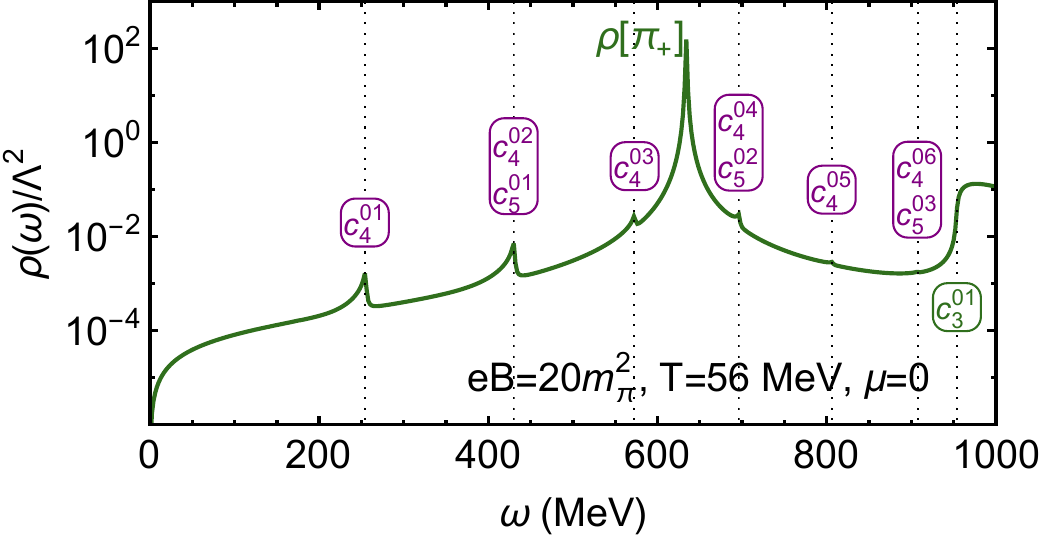}
\caption{Spectral function of $\pi_+$ at $T=56$MeV, $\mu=0$ and magnetic field $eB=20m_\pi^2$. Inserted annotations refer to the different processes, $c_3^{ij}:\pi_+'\rightarrow u_i \bar{d}_j$, $c_4^{ij}:\pi_+'\bar{u}_i \rightarrow \bar{d}_j$, $c_5^{ij}:\pi_+'{d}_i\rightarrow u_j $.}
\label{spectral_T56_B20}
\end{figure}
We further provide the $\pi_+$ spectral function at $T = 56\ \mathrm{MeV}$, $\mu = 0$, and $eB = 20\ m_\pi^2$ in the appendix. This comparison is made with the corresponding spectral function at $\mu = 200\ \mathrm{MeV}$ shown in the top panel of Fig.~\ref{spectral_mu}. At vanishing chemical potential, the annihilation processes $\pi_+' \bar{u}i \to \bar{d}j$ and $\pi+' d_i \to u_j$ contribute equally. In contrast, at $\mu = 200\ \mathrm{MeV}$, the channel $\pi+' d_i \to u_j$ is noticeably enhanced, while $\pi_+' \bar{u}_i \to \bar{d}_j$ becomes highly suppressed.

\section{threshold function}
\label{threshold}
With the boson and fermion occupation numbers and their derivatives,
\begin{eqnarray}
&& n_B(x)=\frac{1}{e^{x/T}-1},\quad n_F(x)=\frac{1}{e^{x/T}+1},\nonumber\\
&& n'_B(x)=\frac{dn_B(x)}{d x},\quad n'_F(x)=\frac{dn_F(x)}{d x},
\end{eqnarray}
the threshold functions $I_\phi^{(1)T}$ and $I_f^{(1)T}$ in the flow equation for effective potential and $I_\phi^{(2)T}$ with four-line vertices are explicitly expressed as
\begin{eqnarray}
I_\phi^{(1)T}(E_{\phi})&=&T\sum_{q_0}\frac{1}{(q_0^2+E_\phi^2)}~=~\frac{1+2n_B(E_{\phi})}{2E_{\phi}},\nonumber\\
I_f^{(1)T}(E_f)&=&T\sum_{q_0}\frac{1}{((q_0-i \mu )^2+E_f^2)}~=~\frac{1-n_F(E_f-\mu)-n_F(E_f+\mu)}{2E_f},\nonumber\\
I^{(2)T}_{\alpha}(E_{\alpha})&=&T\sum_{q_0}\frac{1}{(q_0^2+E_\alpha^2)^2}~=~\frac{1}{4}\Big(\frac{1+2n_B(E_\alpha)}{E_\alpha^3}-\frac{2n_B'(E_\alpha)}{E_\alpha^2}\Big).
\end{eqnarray}
After a straightforward but tedious calculation, the threshold function $L_{\phi_i\phi_j}(p_0)$ for meson loops with three-line vertices can be written as
\begin{eqnarray}
\label{thresholdJmeson}
J_{\alpha\alpha}^T&=&T\sum_{q_0}\frac{1}{(q_0^2+E_\alpha^2)^2((q_0+p_0)^2+E_\alpha^2)}\nonumber\\
&=&\frac{12E_{\alpha}^2+p_0^2}{4E_{\alpha}^3(4E_{\alpha}^2+p_0^2)^2}(1+2n_B(E_{\alpha}))-\frac{1}{2E_{\alpha}^2(4E_{\alpha}^2+p_0^2)}n'_B(E_\alpha),\nonumber\\
J_{\alpha\beta}^T&=&T\sum_{q_0}\frac{1}{(q_0^2+E_\alpha^2)^2((q_0+p_0)^2+E_\beta^2)}\nonumber\\
&=&\frac{ (E_{\alpha }^2+E_{\beta }^2-(2 E_{\alpha }-i
   p_0)^2)}{4 E_{\alpha }^3 (E_{\beta }^2-(E_{\alpha }-i p_0)^2)^2}(1+n_B(E_{\alpha }))
   +\frac{(E_{\alpha }^2+E_{\beta }^2-(2 E_{\alpha }+i p_0)^2) }{4
   E_{\alpha }^3 (E_{\beta }^2-(E_{\alpha }+i p_0)^2)^2}n_B(E_{\alpha })\nonumber\\
   &&+\frac{n'_B(E_{\alpha })}{4E_{\alpha }^2 ((E_{\alpha }-i p_0)^2-E_{\beta }^2)}
   +\frac{n'_B(E_{\alpha })}{4E_{\alpha }^2 ((E_{\alpha }+i p_0)^2-E_{\beta }^2)}\nonumber\\
   &&+\frac{1}{2 E_{\beta } (E_{\alpha }^2-(E_{\beta }+i p_0)^2)^2}(1+n_B(E_{\beta }))
   +\frac{1}{2 E_{\beta } (E_{\alpha }^2-(E_{\beta }-i p_0)^2)^2}n_B(E_{\beta }).
   \end{eqnarray}
For fermion loops, we divide ${J}_{ff}^{\sigma T}(p_0,\mu)$ into two terms for convenience, 
\begin{eqnarray}
{J}_{ff}^{\sigma T}(p_0,\mu)&=&{J}_f^{(0)T}(p_0,\mu)+4m^2{J}_f^{(1)T}(p_0,\mu),\nonumber\\
{J}_{ff}^{\pi_0 T}(p_0,\mu)&=&{J}_f^{(0)T}(p_0,\mu).
\end{eqnarray}
Threshold functions ${J}_{ff}^{(0,1)T}(p_0,\mu)$ encoding the Matsubata sum of loops composed by same flavor of quark are given by
\begin{eqnarray}
{J}_{ff}^{(0)T}(p_0,\mu)&=&\frac{1}{2}T\sum_{q_0}\frac{-E_f^2-(q_0-i \mu ) (q_0 +2 p_0-i \mu)}{((q_0-i \mu )^2+E_f^2)^2 ((q_0+p_0-i \mu)^2+E_f^2)}+\{\mu\leftrightarrow -\mu\}\nonumber\\
&=&\frac{(4 E_f^2-p_0^2) }{E_f (4 E_f^2+p_0^2)^2}(1-n_F(E_f-\mu)-n_F(E_f+\mu))+\frac{1}{4 E_f^2+p_0^2}(n'_F(E_f-\mu )+n'_F(E_f+\mu )),\nonumber\\
{J}_f^{(1)T}(p_0,\mu)&=&\frac{1}{2}T\sum_{q_0}\frac{1}{((q_0-i \mu )^2+E_f^2)^2 ((q_0+p_0-i \mu)^2+E_f^2)}+\{\mu\leftrightarrow -\mu\}\nonumber\\
&=&-\frac{(12E_f^2+p_0^2) }{4 E_f^3 (4 E_f^2+p_0^2)^2}(1-n_F(E_f-\mu )-n_F(E_f+\mu ))-\frac{1}{4E_f^2(4 E_f^2+p_0^2)}(n'_F(E_f-\mu)+n'_F(E_f+\mu)).\nonumber
\end{eqnarray}
Threshold function $J_{du,mn}^{(0,1)}(p_0,\mu)$ with respect to the loop composed by a u-quark and a d-quark is given by 
\begin{eqnarray}
J_{du,mn}^{(0)}(p_0,\mu)&=&T \sum_{q_0}\frac{-(q_0-i\mu)(q_0+2p_0-i\mu)-E_{d,m}^2}{[{(q_0+p_0-i\mu)^2+E_{u,n}^2}][{(q_0-i\mu)^2+E_{d,m}^2}]^2}\nonumber\\
&=&-\frac{(E_{d,m}^2-E_{u,n}^2+p_0^2) }{2 E_{d,m} (E_{u,n}^2-(E_{d,m}-i p_0)^2)^2}(1-n_F(E_{d,m}+\mu ))
+\frac{(E_{d,m}^2-E_{u,n}^2+p_0^2) }{2 E_{d,m}(E_{u,n}^2-(E_{d,m}+i p_0)^2)^2}n_F(E_{d,m}-\mu )\nonumber\\
&&
-\frac{(-E_{d,m}^2+E_{u,n}^2+p_0^2)}{2 E_{u,n} (E_{d,m}^2-(E_{u,n}+i p_0)^2)^2}(1-n_F(E_{u,n}+\mu ))
+\frac{(-E_{d,m}^2+E_{u,n}^2+p_0^2) }{2 E_{u,n} (E_{d,m}^2-(E_{u,n}-i p_0)^2)^2}n_F(E_{u,n}-\mu )\nonumber\\
&&
+\frac{i p_0 }{2 E_{d,m}(E_{u,n}^2-(E_{d,m}-i p_0)^2)}n'_F(E_{d,m}+\mu)
-\frac{i p_0 }{2 E_{d,m} (E_{u,n}^2-(E_{d,m}+i p_0)^2)}n'_F(E_{d,m}-\mu),
\nonumber\\
J_{du,mn}^{(1)}(p_0,\mu)&=&T \sum_{q_0}\frac{1}{[{(q_0+p_0-i\mu)^2+E_{u,n}^2}][{(q_0-i\mu)^2+E_{d,m}^2}]^2}\nonumber\\
&=&-\frac{(E_{d,m}^2+E_{u,n}^2-(2 E_{d,m}-i p_0)^2) }{4 E_{d,m}^3 (E_{u,n}^2-(E_{d,m}-i p_0)^2)^2}(1-n_F(E_{d,m}+\mu ))
+\frac{(E_{d,m}^2+E_{u,n}^2-(2 E_{d,m}+i p_0)^2) }{4 E_{d,m}^3 (E_{u,n}^2-(E_{d,m}+i p_0)^2)^2}n_F(E_{d,m}-\mu )\nonumber\\
&&
-\frac{1-n_F(E_{u,n}+\mu )}{2 E_{u,n} (E_{d,m}^2-(E_{u,n}+i p_0)^2)^2}
+\frac{n_F(E_{u,n}-\mu )}{2 E_{u,n} (E_{d,m}^2-(E_{u,n}-i p_0)^2)^2}\nonumber\\
&&
+\frac{n'_F(E_{d,m}+\mu )}{4 E_{d,m}^2 (-E_{u,n}^2+(E_{d,m}-i p_0)^2)}
+\frac{n'_F(E_{d,m}-\mu )}{4 E_{d,m}^2 (-E_{u,n}^2+(E_{d,m}+i p_0)^2)}.
\end{eqnarray}
Likewise, with the regulator rest of the other propagator, one has $J_{ud,nm}^{(0,1)}(p_0,\mu)$ from the above by exchanging $E_{u,n}\leftrightarrow E_{d,m}$,
\begin{eqnarray}
J_{ud,nm}^{(0)}(p_0,\mu)&=&T \sum_{q_0}\frac{-(q_0-i\mu)(q_0+2p_0-i\mu)-E_{u,n}^2}{[{(q_0+p_0-i\mu)^2+E_{d,m}^2}][{(q_0-i\mu)^2+E_{u,n}^2}]^2}~=~J_{du,mn}^{0}(p_0,\mu)\{E_{u,n}\leftrightarrow E_{d,m}\},\nonumber\\
J_{ud,nm}^{(1)}(p_0,\mu)&=&T \sum_{q_0}\frac{1}{[{(q_0+p_0-i\mu)^2+E_{d,m}^2}][{(q_0-i\mu)^2+E_{u,n}^2}]^2}~=~J_{du,mn}^{1}(p_0,\mu)\{E_{u,n}\leftrightarrow E_{d,m}\}.
\end{eqnarray}

\section{Meson loop and quark loop with different electric charge}
\label{MloopFloop}
Translating from the Feynman digram, the meson loop composed by $\sigma$ and $\pi_+$ with two three-line vertex is 
\begin{eqnarray}
\label{chargedloop}
\text{M-Loop}&=&(\Gamma^{(3)}_{\sigma\pi\pi})^2\int_{r,r'}\phi_{np}({r'})G_{\pi_+}(r,r')G_{\sigma}(r',r)\phi^*_{np}({r})\nonumber\\
&=&(\Gamma^{(3)}_{\sigma\pi\pi})^2\int d^4 r d^4 r' e^{-ip_0 t'+ip_3z'}\Psi_{np_1}(\vec{r}'_\perp)e^{ip_0 t-ip_3z}\Psi_{np_1}^*(\vec{r}_\perp)\nonumber\\
&&\times  \int\frac{d^4qd^4k}{((2\pi)^4)^2} e^{-i(k-q)\cdot(r-r')}e^{-is_\perp\frac{(x-x')(y+y')}{2l^2}}{G}_{\pi_+}(k) {G}_{\sigma}(q),
\end{eqnarray}
where $r=(t,x,y,x)$, $r'=(t',x',y',x')$.  $G_{\pi_+}(r,r')$ and $G_{\sigma}(r',r)$ are propagators of $\pi_+$ and $\sigma$ in coordinate space respectively, they are the Fourier transformation of propagators in momentum space. For $\pi_+$, it is given by \eqref{picpropagator}. The integral over $t$ and $z$ yields the momentum conservation in the zeroth and third component, while the integral over $\vec{r}_\perp$ and $\vec{r}'_\perp$ is nontrivial, and can be carried out using the following expression,  
\begin{eqnarray}
\int_{-\infty}^{+\infty}e^{ixy}e^{-x^2/2}H_n(x)dx=\sqrt{2\pi} i^n e^{-y^2/2}H_n(y).
\end{eqnarray} 
Considering that we have set the external line to be in the ground state $n=0$, the integral over coordinates in \eqref{chargedloop} yields
\begin{eqnarray}
\label{chargedloop1}
\text{M-Loop}=(\Gamma^{(3)}_{\sigma\pi\pi})^24l^2\sum_{m=0}^{\infty}\int\frac{d^2k_\perp}{(2\pi)^2}\frac{d^4q}{(2\pi)^4} e^{-2k^2_\perp l^2}e^{-q_\perp l^2}e^{-2l^2\vec{q}_\perp\cdot\vec{k}_\perp}\frac{1}{q_0^2-\vec{q}^2-m_\sigma^2}\frac{(-1)^mL_m(2k^2_\perp l^2)}{(q_0+p_0)^2-(2m+1)|eB|-q_3^2-m_\pi^2}.
\end{eqnarray}
The integral over $\vec{k}_\perp$ can then be done using 
\begin{eqnarray}
\int_0^{\infty} x e^{-x^2}L_n(x^2)J_0(x y)=\frac{2^{-2n-1}}{n!}y^{2n}e^{-\frac{1}{4}y^2}.
\end{eqnarray} 
Finally, one arrives at 
\begin{eqnarray}
\label{chargedloop2}
\text{M-Loop}&=&(\Gamma^{(3)}_{\sigma\pi\pi})^2
\sum_{m=0}^{\infty}\int\frac{d^4q}{(2\pi)^4}
\frac{1}{q_0^2-\vec{q}^2-m_\sigma^2}\frac{1}{(q_0+p_0)^2-(2m+1)|eB|-q_3^2-m_\pi^2} \frac{1}{m!}\Big(\frac{l^2 q_\perp^2}{2}\Big)^m e^{-\frac{1}{2}l^2q_\perp^2}.
\end{eqnarray} 
One can then implement the Wick rotation and FRG modification by taking the replacement $p_3^2\rightarrow p_3^2(1+r_B)$. The flow equation is then derived by taking derivative respective to RG-scale $k$. Taking the limit $|eB|\rightarrow 0$, one can obviously find that the summation over Landau level becomes trivial. With $(2m+1)|eB|$ becoming $\vec{q}_\perp^2$, the summation $\sum_{m=0}^{\infty}\frac{1}{m!}\Big(\frac{l^2 q_\perp^2}{2}\Big)^m =e^{\frac{1}{2}l^2q_\perp^2}$ cancels with the other exponential in \eqref{chargedloop2}. Then \eqref{chargedloop2} goes back to the expression without the background magnetic field. 

The quark loop constructed by u quark and d quark also appears in the $\pi_+$ two-point function. The Schwinger phase of the u d quark cannot cancel out, thus one has to calculate the quark loop starting from the coordinate space. The quark propagator in the coordinate space can also be obtained by Fourier transformation \eqref{picpropagator} from the momentum space propagator \eqref{udpropagator}. This fermion loop in the coordinate space is 
\begin{eqnarray}
\label{quarkloop}
\text{F-loop}&=&\int_{r,r'}\text{Tr}\Big[\phi_{np}({r'})(\sqrt{2}ig\gamma_5)G_u(r,r')(\sqrt{2}ig\gamma_5)G_d(r',r)\phi^*_{np}({r})\Big]\nonumber\\
&=&2\int d^4 r d^4 r' e^{-ip_0 t'+ip_3z'}\Psi_{np_1}(\vec{r}'_\perp)e^{ip_0 t-ip_3z}\Psi_{np_1}^*(\vec{r}_\perp) \times\int\frac{d^4k}{(2\pi)^4}e^{-ik_0(t-t')+i\vec{k}\cdot(\vec{r}-\vec{r}')}e^{-is_\perp^u\frac{(x-x')(y+y')}{2l_u^2}}\nonumber\\
&&\times\int\frac{d^4q}{(2\pi)^4}e^{-iq_0(t'-t)+i\vec{q}\cdot(\vec{r}'-\vec{r})}e^{-is_\perp^d\frac{(x'-x)(y+y')}{2l_d^2}}\times \text{tr}[{G}_{u}(k)i\gamma^5{G}_{d}(q)i\gamma^5],
\end{eqnarray}
taking $l=1/\sqrt{|eB|}$, then for u-quark $s_\perp^u=+1$, $l_u^2=\frac{3}{2}l^2$, and for d-quark $s_\perp^d=-1$, $l_d^2=3l^2$. The integral over $t,z$ gives momentum conservation in zeroth and third component of momentum, while the integral over $\vec{r}_\perp$ and $\vec{r}'_\perp$ gives
\begin{eqnarray}
\label{integralyy}
\int dxdx' \int dydy' e^{i \vec{k}_\perp\cdot(\vec{r}_\perp-\vec{r}'_\perp)+i \vec{q}_\perp\cdot(\vec{r}'_\perp-\vec{r}_\perp)}e^{-i\frac{(x-x')(y+y')}{2l^2}}\Psi_{np_1}(\vec{r}'_\perp)\Psi_{np_1}^*(\vec{r}_\perp)
&=&{2\pi} 2l^2e^{-l^2(\vec{k}_\perp-\vec{q}_\perp)^2},
\end{eqnarray}
where the external line $\Psi_{np_1}(\vec{r}'_\perp)$ is given by \eqref{externalline}, and for the external line, we take the lowest Landau level $n=0$. After evaluating the Dirac trace, integrating $k_0$ and  $k_3$ using the $\delta-$function and analytically completing the integral over $\vec{k}_\perp$ using  
\begin{eqnarray}
\int_0^\infty x^{\nu+1} e^{-\beta x^2} L_n^\nu(\alpha x^2) J_\nu (x y) dx=2^{-\nu-1}\beta^{-\nu-n-1}(\beta-\alpha)^n y^\nu e^{-\frac{y^2}{4\beta}} L_n^\nu[\frac{\alpha y^2}{4\beta(\alpha-\beta)}],
\end{eqnarray}
 one finally arrives at
\begin{eqnarray}
\label{quarkloop3}
\text{F-loop}&=&4\int\frac{ d^4q}{(2\pi)^4}\sum_{h,m=0}^\infty\frac{(-1)^{m}}{[(p_0+q_0)^2-2h|q_uB|-(p_3+q_3)^2-m_u^2][q_0^2-2m|q_dB|-q_3^2-m_d^2]}\frac{8}{  5^{h}}\nonumber\\
&&\times\Big\{(q_0(p_0+q_0)-q_3(p_3+q_3)-m_um_d)\Big[e^{-\frac{18}{5}q_\perp^2l^2}\Big(\frac{1}{5}L_h(-\frac{12}{5}q_\perp^2l^2) L_{m-1}(6q_\perp^2l^2)-L_{h-1}(-\frac{12}{5}q_\perp^2l^2)L_{m}(6q_\perp^2l^2)\Big)\Big]\nonumber\\
&&\quad-\frac{16}{5}q_\perp^2e^{-\frac{18}{5}q_\perp^2l^2}L_{h-1}^1(-\frac{12}{5}q_\perp^2l^2)L_{m-1}^1(6q_\perp^2l^2)\Big\}.
\end{eqnarray}
The integral over perpendicular momentum $\vec{q}_\perp$ can be performed analytically, giving two functions $Y_1(m,n)$ and $Y_2(m,n)$ defined in \eqref{Y1Y2}. Then we perform the Wick rotation to Euclidean space, consider the RG modification to the propagator and the chemical potential. The quark loop in the two point function becomes 
\begin{eqnarray}
2g^2\widetilde{\partial}_k\text{Tr}[G_u(i\gamma^5)G_d(i\gamma^5)]&=&2g^2N_c\Big(J^{\pi_+}_{du}(p_0,\mu)+J^{\pi_+}_{ud}(-p_0,\mu)\Big),
\end{eqnarray}
with the loop functions $J^{\pi_+}_{du}$ and $J^{\pi_+}_{ud}$ given by \eqref{udloop}.

\section{Numerical stability analysis}
\label{stability_analysis}
In this appendix, we briefly discuss the numerical stability of the phase diagram. In the present work, due to the use of the anisotropic $p_3$-regulator, the transverse momentum integration and the summation over Landau levels are regulated by an upper energy scale $\Lambda_T$. To assess the sensitivity of the phase structure to this cutoff, we compare the phase boundaries obtained at $eB = 20\,m_\pi^2$ for two representative values, $\Lambda_T = 3~\text{GeV}$ and $5~\text{GeV}$.

 \begin{figure}[H]\centering
\includegraphics[height=0.25\textwidth]{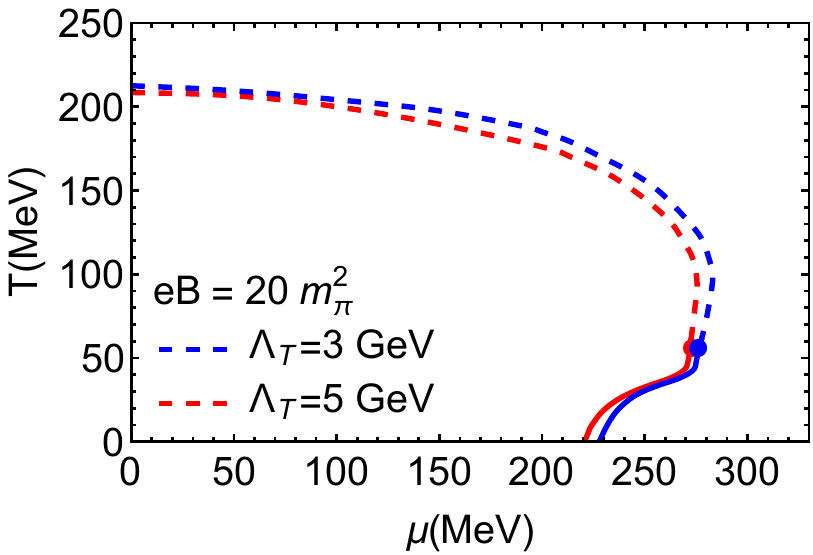}
\caption{Comparison of phase boundary at $eB = 20\,m_\pi^2$ with $\Lambda_T = 3~\text{GeV}$ (blue) and $5~\text{GeV}$ (red).} 
\label{phase2compare}
\end{figure}

The comparison is shown in Fig.~\ref{phase2compare}. We observe that while the precise location of the phase boundary and the critical endpoint shows a mild quantitative dependence on $\Lambda_T$, the overall structure of the phase diagram remains unchanged. In particular, the existence of a first-order transition line, its extension toward lower chemical potential with increasing magnetic field, and the qualitative position of the critical endpoint are robust. We have also verified that increasing the number of grid points in the field discretization does not lead to visible changes in the phase structure.In this work the phase diagram serves primarily as a background input for the computation of meson spectral functions. Our main focus is on the qualitative behavior of spectral functions in different phases, rather than on a high-precision determination of the phase boundary itself. Therefore, the observed level of numerical uncertainty does not affect the main conclusions of the paper.

\section{Origin of oscillatory structures in the $\pi^+$ spectral function}
\label{oscillation_analysis}
 \begin{figure}[H]\centering
\includegraphics[width=0.329\textwidth]{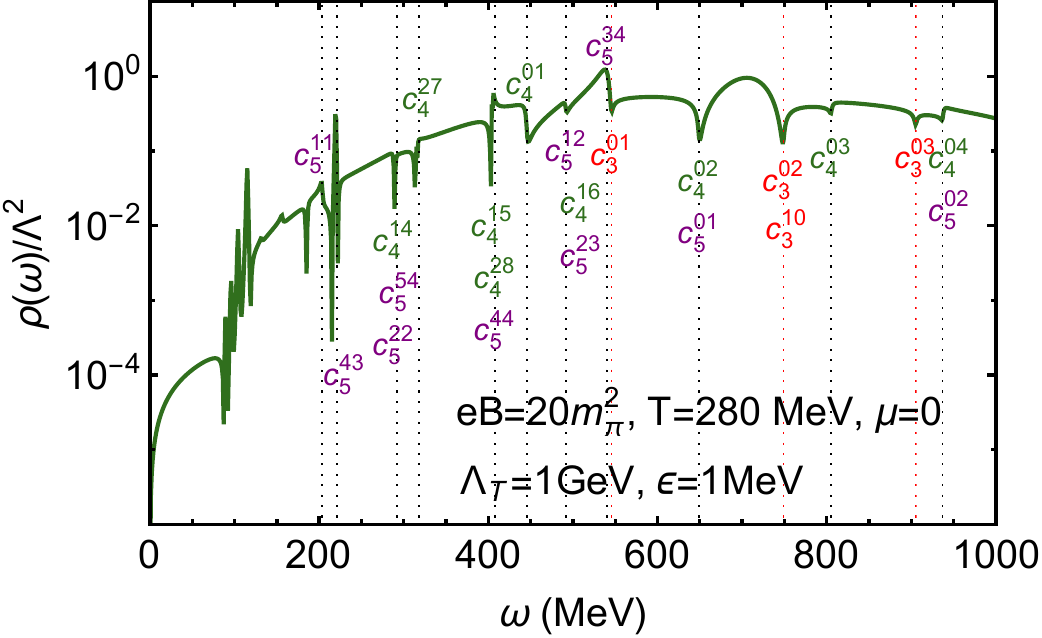}
\includegraphics[width=0.329\textwidth]{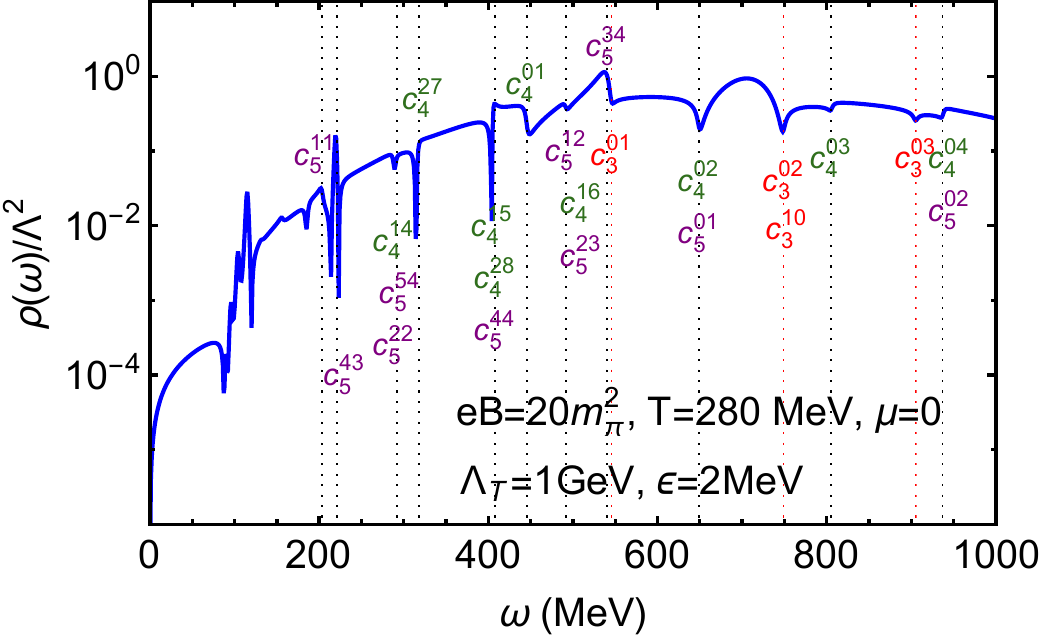}
\includegraphics[width=0.329\textwidth]{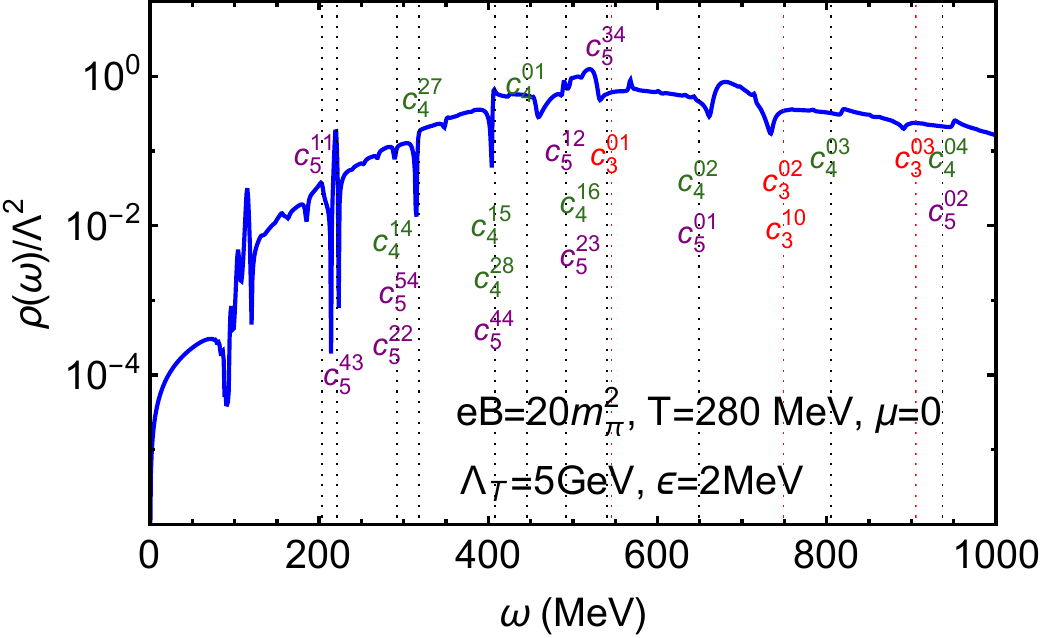}
\caption{Spectral function of $\pi_+$ at $eB = 20\,m_\pi^2$ and $T=280$ MeV with $\Lambda_T = 1~\text{GeV}$ and $\Lambda_T = 5~\text{GeV}$ and with different $\epsilon$.}
\label{app_T280}
\end{figure}
In this appendix, we provide a detailed analysis of the oscillatory structures observed in the $\pi^+$ spectral function at high temperature. As shown in Fig.~4 of the main text, the $\pi^+$ spectral function develops pronounced oscillations at high temperature. These structures originate from the large number of available annihilation and decay channels involving quarks in different Landau levels.

To demonstrate this explicitly, we perform calculations at $eB = 20\,m_\pi^2$ and $T = 280~\text{MeV}$, where the thermal population of higher Landau levels is significant. We compare results obtained with different transverse cutoffs $\Lambda_T$. In Fig.~\ref{app_T280}, we show the spectral functions for $\Lambda_T = 1~\text{GeV}$ and $\Lambda_T = 5~\text{GeV}$. For smaller $\Lambda_T$, fewer Landau levels contribute, and the oscillatory structures are significantly reduced. For larger $\Lambda_T$, more Landau levels are included, leading to a dense sequence of oscillations. This clearly demonstrates that the oscillations are associated with the inclusion of higher Landau levels. We further investigate the dependence on the regulator parameter $\epsilon$ used in the analytic continuation. As shown in Fig.~\ref{app_T280}, a smaller $\epsilon$ leads to sharper oscillations, while increasing $\epsilon$ smooths out these structures. This indicates that the oscillations are related to threshold-like features in the spectral function.

 \begin{figure}[H]\centering
\includegraphics[width=0.45\textwidth]{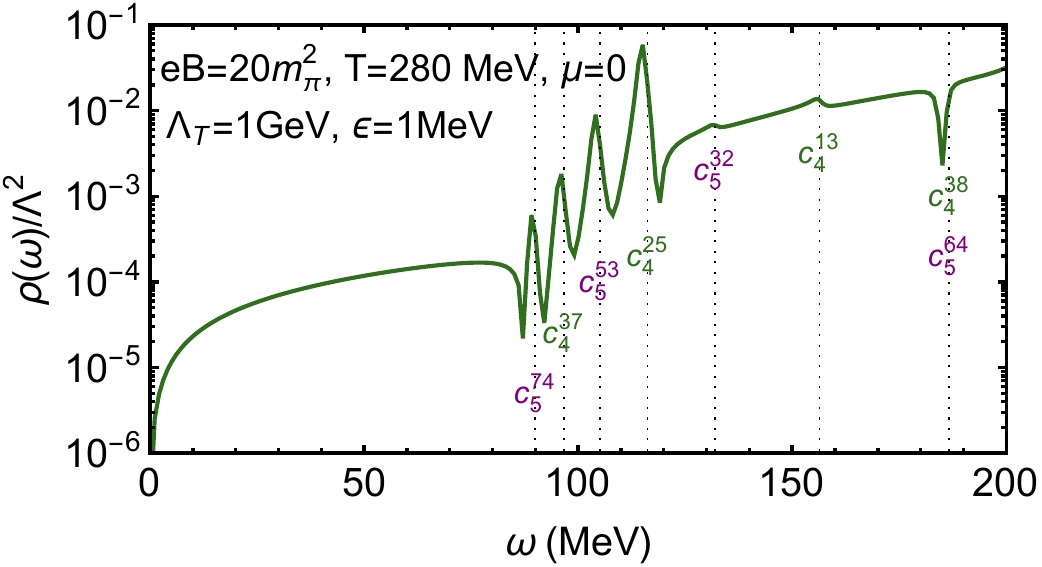}
\includegraphics[width=0.45\textwidth]{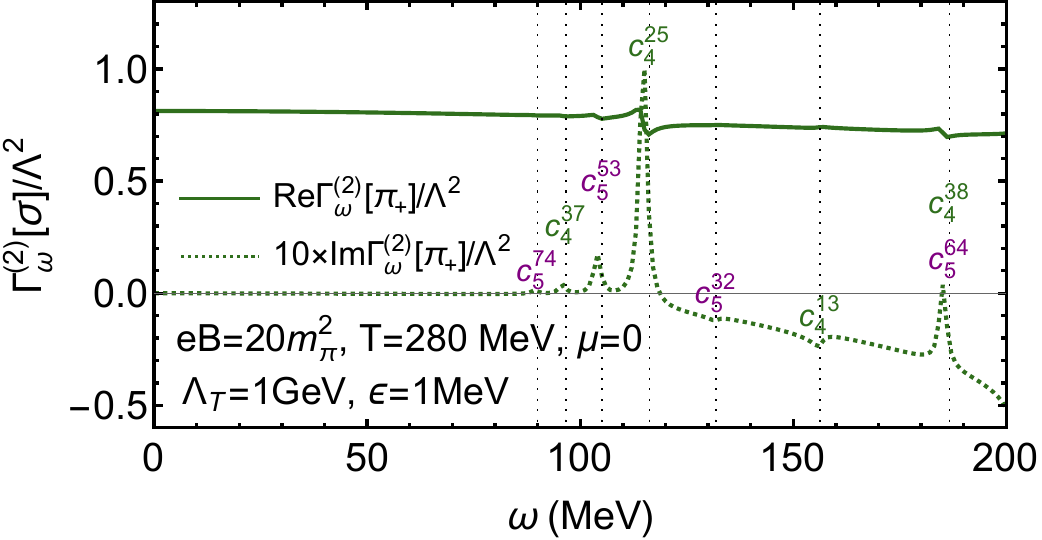}
\caption{Low-energy region of $\pi_+$ spectral function at $eB = 20\,m_\pi^2$ and $T=280$ MeV with $\Lambda_T = 1~\text{GeV}$ and with $\epsilon=1$MeV. Right panel: corresponding real and imaginary part of the two point function, with imaginary part enlarged by 10 times.}
\label{app_T280_zoom}
\end{figure}
To understand their microscopic origin, we analyze the contributions from individual channels. In Fig.~\ref{app_T280_zoom}, we focus on the low-energy region, where the oscillations are most pronounced. Each peak can be identified with a specific process involving quarks in different Landau levels. The relevant processes can be classified as follows: $c_3^{ij}: \pi_+' \rightarrow u_i \bar{d}_j$ (decay channel), $c_4^{ij}: \pi_+' \bar{u}_i \rightarrow \bar{d}_j$ (annihilation with thermal antiquark), $c_5^{ij}: \pi_+' d_i \rightarrow u_j$ (annihilation with thermal quark). The correspondence between these channels and the oscillatory structures is further confirmed by analyzing the real and imaginary parts of the two-point function, shown in the right panel of Fig.~\ref{app_T280_zoom}. We observe that each oscillation is associated with a rapid variation
in $\text{Im}\,\Gamma^{(2)}_\omega$, accompanied by a kink-like structure in $\text{Re}\,\Gamma^{(2)}_\omega$, characteristic of threshold openings.

These results demonstrate that the oscillatory structures in the $\pi^+$ spectral function are physical in origin and arise from the rich set of Landau-level-resolved scattering and decay processes at high temperature.

\bibliographystyle{unsrt} 

\bibliography{ref}

\end{document}